# Plasmonic photoconductive terahertz focal-plane array with pixel super-resolution

**Authors:** Xurong Li[1,2] †, Deniz Mengu[1,2,3] †, Aydogan Ozcan[1,2,3] and Mona Jarrahi[1,2,*]

**Affiliations:**

[1]Department of Electrical & Computer Engineering, University of California Los Angeles (UCLA), California, USA

[2]California NanoSystems Institute (CNSI), University of California Los Angeles (UCLA), California, USA

[3]Department of Bioengineering, University of California Los Angeles (UCLA), California, USA

*Corresponding author. Email: mjarrahi@ucla.edu

† Equal contribution

## Abstract

Imaging systems operating in the terahertz part of the electromagnetic spectrum are in great demand because of the distinct characteristics of terahertz waves in penetrating many optically-opaque materials and providing unique spectral signatures of various chemicals. However, the use of terahertz imagers in real-world applications has been limited by the slow speed, large size, high cost, and complexity of the existing imaging systems. These limitations are mainly imposed due to the lack of terahertz focal-plane arrays (THz-FPAs) that can directly provide the frequency-resolved and/or time-resolved spatial information of the imaged objects. Here, we report the first THz-FPA that can directly provide the spatial amplitude and phase distributions, along with the ultrafast temporal and spectral information of an imaged object. It consists of a two-dimensional array of ~0.3 million plasmonic photoconductive nanoantennas optimized to rapidly detect broadband terahertz radiation with a high signal-to-noise ratio. As the first proof-of-concept, we utilized the multispectral nature of the amplitude and phase data captured by these plasmonic nanoantennas to realize pixel super-resolution imaging of objects. We successfully imaged and super-resolved etched patterns in a silicon substrate and reconstructed both the shape and depth of these structures with an effective number of pixels that exceeds 1-kilo pixels. By eliminating the need for raster scanning and spatial terahertz modulation, our THz-FPA offers more than a 1000-fold increase in the imaging speed compared to the state-of-the-art. Beyond this proof-of-concept super-resolution demonstration, the unique capabilities enabled by our plasmonic photoconductive THz-FPA offer transformative advances in a broad range of applications that use hyperspectral and three-dimensional terahertz images of objects for e.g., industrial inspection, security screening, and medical diagnosis, among others.

## Introduction

The past decades witnessed significant developments in terahertz imaging technologies and their unique utilization in a broad range of applications. The relatively low energy of terahertz photons and their high penetration through many non-conductive materials make terahertz radiation promising for non-destructive biomedical imaging[1–3], security screening[4–6], quality control of pharmaceutical[7–9], industrial[10–12], and agricultural products[13–15], as well as cultural heritage conservation[16–18].

Despite its great potential, the usage of terahertz imaging in real-world applications has been limited by the slow speed, large size, high cost, and complexity of the existing imaging systems. Terahertz time-domain spectroscopy (THz-TDS) systems have been instrumental for various imaging applications[19–29] since they provide the time-resolved response of objects with a sub-picosecond temporal resolution, which yields both amplitude and phase information over a broad terahertz frequency range. However, due to the single-pixel nature of existing THz-TDS systems, image data are acquired by raster-scanning either the object or the imaging system, leading to very slow imaging speeds and bulky, complex setups.

To avoid raster-scanning, electro-optic processes in nonlinear crystals have been employed to convert the object's terahertz amplitude and phase information to the near-infrared regime and acquire them with an optical camera[22–25]. However, due to the nonlinear nature of the wavelength conversion process, these imaging systems generally require bulky and expensive high-energy lasers and provide very low signal-to-noise ratio (SNR) levels (see Supplementary Table S1). Another way to avoid raster-scanning is to encode the terahertz radiation pattern interacting with the object through a time-varying spatial modulator and reconstruct the image using the spatial pattern information[26–30]. This approach enabled faster and more robust terahertz imaging systems by eliminating the mechanical stages used for raster scanning. However, these imaging systems require multiple time intervals to encode the terahertz radiation pattern with different spatial distributions, limiting their speed, especially when acquiring image data over a broad terahertz bandwidth. In addition, the requirement for a time-varying spatial modulator adds to the size, complexity, and cost of the imaging system.

Further advancements in terahertz imaging technology would significantly benefit from terahertz focal-plane arrays (THz-FPAs) that can directly provide the spatial, ultrafast temporal, spectral, amplitude, and phase information of the object simultaneously. Existing detector arrays based on field-effect transistors[31] and microbolometers[32] do not provide time-resolved and frequency-resolved image data and lack phase information. It has been shown that phase information can be recovered through digital holography when using these detector arrays in an interferometric setup[33–35]. However, the scope of such interferometric setups is restricted to objects that possess axially uniform refractive index and does not apply to more general multi-layered structures.

All in all, imaging science in the terahertz part of the electromagnetic spectrum has been lacking FPAs that can directly provide the amplitude and phase information of samples over a large spectral bandwidth and with an ultrafast temporal response. Such capabilities, if made possible in the same FPA, would open up various new applications; for example, difficult-to-see phase-only objects that are weakly scattering could be imaged and sensed in a snapshot, potentially revealing their conformational changes at ultra-fast time scales. As another opportunity, broadband spectral features that are acquired from the objects could also reveal their unique spectral signatures, and when this is combined with the amplitude and phase images and the ultra-fast temporal response of the samples, it could provide unprecedented channels of information for automated three-dimensional (3D) analysis and quantification of samples at high throughput. These powerful information channels, once bundled together, would help us fully exploit all the advantageous features of terahertz waves and significantly improve the resolution and throughput of techniques that are used for imaging, sensing, and inspecting materials, objects, scenes, and chemical processes in the terahertz frequency range.

Motivated by these pressing needs, here we present the first THz-FPA that can directly provide the spatial amplitude and phase distributions, along with the ultrafast temporal and spectral information of an imaged object. It consists of a two-dimensional (2D) array of 283,500 plasmonic nanoantennas[36–39] engineered to detect broadband terahertz radiation with a high SNR when used in a THz-TDS system. To simplify data readout from the THz-FPA, these plasmonic nanoantennas are grouped into 7 × 9 clusters, and the collective response of all the nanoantenna clusters is captured electronically at each temporal point to resolve their time-domain response simultaneously. The amplitude and phase responses of the THz-FPA outputs are extracted from the time-domain data over a 3 THz bandwidth. Using an electronic readout, the THz-FPA outputs at each temporal point are captured in 164 μs, exhibiting a 1000-fold increase in speed compared to the fastest-reported multi-pixel terahertz time-domain imaging system[28], enabling time-domain terahertz video capture at 16 fps (see Supplementary Fig. S8 and Supplementary Video). We utilize the broadband nature of the amplitude and phase data captured from the 283,500 plasmonic nanoantennas to realize pixel super-resolution (PSR) using a convolutional neural network (CNN) trained with deep learning. As a first proof-of-concept, we successfully imaged and super-resolved etched patterns in a high-resistivity silicon substrate and reconstructed both the shape and depth of these patterns with an effective number of pixels that exceeds 1 kilo pixels, digitally increasing the space-bandwidth product of the THz-FPA by 16-fold. This PSR-enhanced THz-FPA would be transformative for building high-throughput and high-resolution imaging and inspection systems that benefit from the unique spectral features of terahertz waves, and would open up a plethora of new applications in biomedical imaging, defense/security screening, agriculture, and quality control, among many others.

**Results**

Figure 1 shows the schematic diagram and the operation principles of the terahertz imaging system. It consists of a THz-FPA composed of a 2D array of 283,500 plasmonic nanoantennas fabricated on a low-temperature-grown GaAs (LT-GaAs) substrate, where the plasmonic nanoantennas serve as photoconductive terahertz detector elements. The geometry of the nanoantennas is chosen to obtain a strong spatial overlap between the received terahertz radiation and optical pump beam to achieve high detection sensitivity over a broad terahertz frequency range[36–39] (see Supplementary Fig. S1). To simplify data readout from the THz-FPA, the plasmonic nanoantennas are grouped into 7 × 9 clusters (i.e., pixels) and the collective response of all the clusters is captured using a custom-made readout circuit packaged with the FPA chip (see Supplementary Fig. S3 and the Methods section). Using this configuration, ultrafast temporal and spectral information of all FPA pixels are resolved with more than a 60 dB SNR and a 3 THz detection bandwidth when used in a THz-TDS setup (see Supplementary Fig. S4 and methods section). Although this clustering reduces the number of physically accessible pixels of the FPA to 63, the distributed nature of the 283,500 plasmonic nanoantennas positioned uniformly across the entire image plane, when combined with their high SNR and large bandwidth, allowed us to perform PSR using a trained CNN, digitally increasing the space-bandwidth product of the THz-FPA by 16-fold and achieving an effective pixel number of >1 kilo pixels. Furthermore, the high sensitivity and broad bandwidth of the THz-FPA enable resolving spectral images of objects over a 2.5 THz bandwidth without using raster scanning or spatial modulation of the terahertz beam (see Supplementary Fig. S6).

As a proof-of-concept, using this unique THz-FPA, we experimentally realized PSR and digitally increased the effective pixel count of our THz-FPA by more than an order of magnitude. For this, we utilized a convolutional deep neural network that was trained based on experimental image data. Figure 2a illustrates the architecture of the convolutional deep neural network used in this study to process the information recorded by the THz-FPA (for details, see Methods section). To train the PSR network with our THz-FPA, we imaged spatially-structured objects comprised of etched patterns in a high-resistivity silicon substrate (see Supplementary Fig. S7). Two different experimental setups were designed and tested to image various patterned objects at a distance of 1.1 mm and 20 cm (far-field) from the THz-FPA (see Supplementary Fig. S5). The same PSR framework was applied to super-resolve the images that are projected onto the THz-FPA in these setups, making our results broadly applicable to different lens-based and lens-free imaging systems.

First, we fabricated 300 spatially structured silicon samples, each corresponding to a unique combination of 75 designed patterns and 4 etch depths ($\Delta_t = 10, 20, 30, and\ 40\mu m$). The objects were imaged at an axial distance of 1.1 mm from the active area of the THz-FPA using a high refractive index semiconductor substrate, leading to a Fresnel number of > 1. Therefore, one of the main sources of spatial resolution loss for these imaging experiments is the pixelation at the THz-FPA. The resulting measurements from these samples were divided into 2 sets for the training and blind testing of our PSR-enhanced THz-FPA. The etched depths of all samples, $\Delta_t^m$, were measured using a profilometer to establish the ground truth images during the training and testing of our PSR deep neural network.

Our PSR neural network aims to benefit from the large SNR and broad bandwidth of each FPA pixel to achieve super-resolution, and therefore it is designed to take the raw spectral amplitude and phase components of the THz-FPA output (Fig. 2a). The spectral amplitude components of each FPA pixel at $N$ different frequencies within $[f_{min}, f_{max}]$ are used as $N$ input channels to the PSR network. We also enriched this input information by concatenating a thickness (i.e., optical path length) image estimate that is computed based on the slope of the unwrapped phase detected within $[f_{min}, f_{max}]$. As a result, we have in total $N + 1$ information channels that feed the PSR network, which was trained using error back-propagation with a structural loss term calculated against the ground truth images of the same samples imaged using a profilometer. This is a one-time training process, which utilizes the high SNR and multispectral output of our THz-FPA to digitally achieve PSR and increase the space bandwidth product of our FPA. Although the detection bandwidth of the THz-FPA extends from 0.1 THz to 3 THz, in our training and testing we took $f_{min}$ and $f_{max}$ as 0.5 THz and 1.0 THz, respectively, avoiding some of the water vapor absorption lines, setting the number of amplitude channels at the input of the PSR network to $N =$

41. As a result of this choice, the PSR network was trained to process the spatial information contained in the $N + 1 = 42$ input channels detected by our THz-FPA to perform PSR, targeting an effective pixel size that is much smaller than the physical pixel size of the THz-FPA.

To train and test the CNN shown in Fig. 2a, we used an experimental dataset of 343 raw image measurements conducted on the 300 samples with some of the samples imaged more than once to demonstrate the repeatability of the imaging system. The samples with spatial features larger than the physical pixel size of the FPA, i.e., 240 $\mu m$ × 270 $\mu m$, were imaged only once, while the other samples that contained subpixel features were imaged more than once at different orientations with respect to the active area of the FPA. This image data was partitioned in a way that 68 measurements were preserved for the blind testing of the PSR network while the rest of the measurements were used for the training (see Supplementary Fig. S15 for the ground truth thickness images of the training data set). Figure 2b depicts the ground truth images of these 68 test samples representing our blind testing set that was never used or seen by the PSR network before, and the corresponding super-resolved images are reported in Fig. 2c. Visual comparison of the ground truth thickness images and the super-resolved PSR network output images reconstructed by taking advantage of the broad bandwidth operation of our multispectral THz-FPA reveals a close match for all the test samples regardless of their material thickness and etch depth. These results also demonstrate resolving small-scale target objects, including patterns with spatial features as small as $1/4^{th}$ of the physical pixel size of our THz-FPA in each lateral direction, which clearly highlights the success of the PSR network. We also performed a quantitative comparison between the ground truth thickness images shown in Fig. 2b and the neural network's reconstructions in Fig. 2c using the structural similarity index measure (SSIM) and the peak signal-to-noise ratio (PSNR); our reconstruction results achieved an SSIM of 0.839±0.098 and a PSNR of 16.60±3.67 dB, further revealing the success of our PSR-enhanced THz-FPA.

To characterize the experimentally achieved resolution of our PSR-enhanced THz-FPA, Fig. 3 provides a closer look at the reconstruction performance of the neural network for some grating patterns in our test data. Specifically, Fig. 3 illustrates different patterns for both vertical and horizontal grating structures with sub-pixel linewidths corresponding to 0.75×, 0.5×, and 0.25× of the physical pixel size of the FPA. The cross-sections of the ground truth and PSR network output images with different sub-pixel linewidths shown in Fig. 3 demonstrate that we can resolve horizontally- and vertically-oriented thickness variations as small as $1/4^{th}$ of the physical pixel size of the THz-FPA. This indicates a 16-fold increase in the number of effective pixels and a super-resolved pixel count of 1008 in total. Furthermore, this PSR performance generalizes over objects with different material thickness variations. For instance, both the horizontal grating in the $2^{nd}$ row and the vertical grating in the $5^{th}$ row of Fig. 3 contain patterns with linewidths that are half of the physical pixel size in the corresponding directions; for these two gratings, the material thickness differences between the etched lines and the surrounding flat region are 40 $\mu m$ and 30 $\mu m$, which correspond to $\lambda_m/10.0$ and $\lambda_m/13.3$, respectively, where $\lambda_m$ is the median illumination wavelength. Despite the deeply sub-wavelength etch depths of these gratings, the PSR network resolved the 2D structure of these sub-pixel gratings and accurately quantified the thickness variations as shown in the cross-sectional plots reported in Fig. 3.

To shed more light on the generalization performance of our PSR-enhanced THz-FPA, we trained 3 additional PSR networks based on different partitioning of the experimental data between the training and testing sets. These new tests and their corresponding super-resolved images are illustrated in Supplementary Fig. S9, all of which successfully demonstrate the generalization capacity of our PSR approach under different training and testing sets.

To further quantify the imaging performance of our PSR-enhanced THz-FPA, in Fig. 4a we report the depth estimation, $\widehat{\Delta_t}$, for each of the reconstructed test patterns shown in Fig. 2b, compared against the ground truth values, $\Delta_t^m$, measured with a profilometer. As desired, $\widehat{\Delta_t}$ inferred by the PSR network output images generally follow the $\Delta_t^m = \widehat{\Delta_t}$ line (dashed line in Fig. 4a). Figures 4b-d illustrate a similar analysis for the 3 additional test datasets shown in Supplementary Fig. S9, where the inference results also follow the $\Delta_t^m = \widehat{\Delta_t}$ line (dashed lines in Figs. 4b-d) without any major errors or outliers, demonstrating the success of the presented PSR-enhanced THz-FPA in achieving both super-resolution and material thickness estimation, independent of the training and testing datasets. This conclusion is also supported by the SSIM and PSNR values shown in Figs. 4e and 4g. SSIM values remain high and vary between 0.839±0.098 and 0.792±0.117 depending on the training/testing partition of the data, and the PSNR values always stay above 15.02±4.80 dB for all of the test datasets, which once again illustrate the generalization success of our PSR-enhanced THz-FPA. We also analyzed the SSIM and PSNR of the reconstructed images across all the test sets using the percentage of thickness estimation error, defined as $\frac{|\Delta_t^m - \widehat{\Delta_t}|}{\Delta_t^m} \times 100$, as shown in Fig. 4f. Based on this analysis, the mean error in $\widehat{\Delta_t}$ inferred by our PSR network is found to be 10.73%; for example for the test samples fabricated with $\Delta_t = 10\ \mu m$, this mean error corresponds to ~1 $\mu m$, which is deeply sub-wavelength.

To further assess the generalization performance of our PSR-enhanced THz-FPA, we designed, fabricated, and imaged various objects with more complex and irregular spatial structures than the ones used for the training, including hand-written numbers and letters, and verified the robustness of our PSR algorithm in resolving both the depth and shape of these objects. These newly fabricated objects were blindly tested with the same PSR neural network used in Fig. 2, and no patterns with similar shapes were seen by the neural network during its training. The depth and shape of the patterns were successfully reconstructed with an average PSNR of 12 dB and SSIM of 0.64, confirming the generalization performance of the presented PSR-enhanced THz-FPA (also see Supplementary Fig. S11). In addition, we numerically tested the same PSR-enhanced THz-FPA framework for resolving images of objects with non-binary depths and weaker contrast. These new objects with variable depths/thicknesses and their corresponding super-resolved images are illustrated in Supplementary Fig. S16; our results reveal the successful reconstruction of these test object images with an average PSNR of 29.4 dB and SSIM of 0.93.

To demonstrate the capability of our PSR-enhanced THz-FPA for imaging far-field objects, we imaged several spatially structured silicon samples with 4 different etch depths ($\Delta_t = 10, 20, 30, and\ 40 \mu m$), placed at an axial distance of 20 cm from the THz-FPA surface (see Supplementary Fig. S5). We constructed an experimental image dataset containing 151 far-field measurements, selected 16 measurements corresponding to 16 unique samples for blind testing, and used the remaining 135 measurements to train the PSR network (see Supplementary Fig. S20 for the ground truth thickness images of the training data set). In our training and testing, we took $f_{min}$ and $f_{max}$ as 1.3 THz and 2.1 THz, respectively, avoiding some of the water vapor absorption lines, setting the number of the amplitude channels at the input of the PSR network to $N = 52$. As a result of this, the PSR network was trained to process the spatial information contained in $N + 1$ = 53 input channels detected by our THz-FPA to perform PSR. The ground truth images of the 16 test samples representing our blind testing set and the corresponding super-resolved images are shown in Figs. 5a and 5b, respectively. Visual comparison of the ground truth thickness images and the super-resolved images indicates a close match for all the test samples regardless of their thickness and etch depth. Our reconstruction results achieved an average SSIM of 0.78 and a PSNR of 14.5 dB, demonstrating the success of our PSR-enhanced THz-FPA imaging system in far-field measurements.

Finally, to experimentally assess the resilience of our PSR-enhanced THz-FPA imaging system to potential object misalignments, we imaged test objects (1) at different axial distances from the THz-FPA and (2) with different rotation angles within the sample field-of-view and demonstrated the successful reconstruction of both the shape and depth of these purposely-misaligned objects despite the fact that the network training was free from such large degrees of misalignments (see Supplementary Figs. S14, S18, and S19).

**Discussion**

In this work, we present the first THz-FPA comprised of 283,500 plasmonic photoconductive nanoantennas, capable of simultaneously providing the spatial amplitude and phase distributions, ultrafast temporal and spectral information of an imaged object. Our PSR-enhanced THz-FPA exploits its multispectral operation to increase the spatial resolution and the effective number of pixels in the final reconstructions compared to the physical pixel size of the FPA. The presented deep learning-driven terahertz imaging system uses the amplitude and phase information of the detected spectral components of an imaged object to enhance the spatial resolution by 4-fold in each lateral direction, increasing the space-bandwidth product of the FPA by 16-fold (4×4), with >1 kilo pixels in each reconstructed image. In addition, our PSR-enhanced THz-FPA has been shown to quantify deeply subwavelength material thickness variations as small as $10\ \mu m$, with a mean thickness accuracy of ~$1\ \mu m$.

To further emphasize the importance of the multispectral operation of the THz-FPA, we compared the performance of our PSR framework with multispectral input data against the use of single-frequency input data. For this purpose, we trained 6 image reconstruction deep neural networks that use the amplitude and phase outputs of the THz-FPA at a single frequency of 0.5, 0.75, 1.0, 1.25, 1.6, and 2.0 THz. The reconstructed images of each one of these single-frequency PSR networks along with their corresponding SSIM and PSNR values are illustrated in Supplementary Fig. S10. While these single-frequency systems can reveal the images of the input objects to some extent, the reconstructed images suffer from severe artifacts and distortions compared to our multispectral PSR results presented in Fig. 2. These results testify that a single-frequency or narrowband THz-FPA capable of providing both amplitude and phase information cannot match the super-resolution capabilities offered by our multispectral THz-FPA. It should be also noted that the existing detector arrays based on field-effect transistors[31] and microbolometers[32], do not provide frequency-resolved image data and cannot obtain phase information, which would prevent their performance enhancement with the discussed PSR framework.

The quality of the resolved images using the reported PSR framework could be further improved with additional training data and unique objects, which would also allow us to make use of deeper neural network architectures without overfitting. Additionally, as outlined in the Methods section, we did not use a generative model trained through an adversarial game between a generator and a discriminator network; instead, we relied solely on a structural loss function in the form of mean-squared-error (MSE). With sufficiently large training data and carefully balanced loss terms using a linear mix between a structural and adversarial loss, the image reconstruction quality reported in this study could be improved further using e.g., Generative Adversarial Networks (GANs)[40,41].

In order to practice our neural network-based PSR approach on new sample types or new imaging systems (with different hardware configurations, image magnification, numerical aperture, etc.) that are not part of the training process, fresh application of the presented learning-based framework (detailed in the Methods section) is recommended for getting optimal results since the training process might introduce a bias toward the spatial features that are well represented in its training dataset, potentially limiting its inference for new types of samples never seen before. Transfer learning from a previously trained network for another type of new sample might speed up the convergence of this learning process. Having emphasized this point, our neural network-based PSR results demonstrated strong generalization to new imaging conditions (e.g., object rotations, axial defocusing) and new types of samples never used in the training before; see, for example, Supplementary Figs. S11, S14, S16, S18, and S19.

The presented multispectral super-resolution framework used for imaging phase objects demonstrates one potential application of our plasmonic photoconductive THz-FPA as its proof-of-concept. To demonstrate that the imaging capabilities of this THz-FPA are not limited to super-resolving phase-only objects through a neural network-based deep learning algorithm, we also captured and super-resolved a video (at 16 fps) of water flow in three adjacent plastic pipes (each with an inner diameter of 250 μm) by using a completely different algorithm based on holography (see Supplementary Fig. S21 and Supplementary Video). Since the water strongly absorbs terahertz waves, the channels that carry the water form an opaque object that is dynamically evolving through the flow, in contrast to earlier transmissive phase objects that were used in our experiments.

Based on all the proof-of-concept experimental results reported in this work, we believe that the unique capabilities of our plasmonic photoconductive THz-FPA enable high-speed acquisition of the spatial, ultrafast temporal, spectral, amplitude, and phase information of an imaged object; this diverse set of information acquired through our THz-FPA could be used in various applications by e.g., utilizing the time-of-flight terahertz pulses to reconstruct 3D images of multi-layered objects[42], using the spectroscopic signatures to identify chemicals[19–21], and integrating with diffractive optical networks for feature detection and object classification[43,44].

The demonstrated THz-FPA can be further advanced to provide a larger number of pixels, larger field-of-view, higher SNR, higher bandwidth, and faster image acquisition. The use of photoconductive nanoantennas integrated with a plasmonic cavity would significantly enhance the spatial overlap between the received terahertz radiation and optical pump beam, offering considerably higher SNR and bandwidth while requiring a much lower optical power level[45]. As a result, the number of nanoantennas forming the THz-FPA pixels can be significantly increased without any degradation in the image quality, enabling larger pixel count FPAs and larger field-of-view imaging systems. Despite offering a 1000-fold increase in the data acquisition speed compared to the fastest-reported multi-pixel terahertz time-domain imaging system[31], the speed of the demonstrated THz-FPA is still limited by the utilized readout electronic system that acquires data from the FPA outputs sequentially. Integration of the FPA with 2D readout integrated circuits (ROICs) would significantly increase the imaging speed, SNR, and bandwidth through a parallel acquisition of data from the FPA outputs. These advancements could bring us much closer to realizing the untapped potential of the terahertz spectrum for numerous applications.

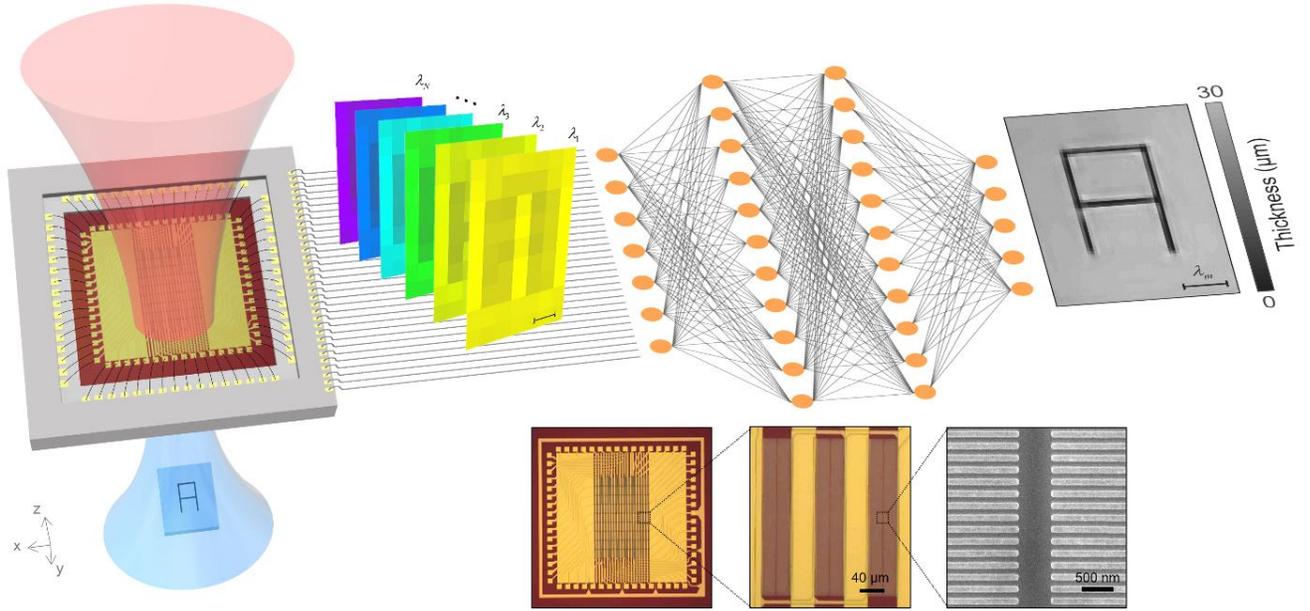

**Figure 1 Plasmonic photoconductive THz-FPA with pixel super-resolution.** The FPA captures hyperspectral, time-resolved terahertz images of an object and a PSR neural network processes these images to reconstruct a higher-resolution image with a 16-fold larger number of effective pixels. The red and blue beams represent the optical pump and terahertz beams, respectively. $\lambda_m = 398$ μm denotes the median wavelength of the terahertz frequency band of interest. Inset shows optical microscope images (left and middle) of the THz-FPA as well as a scanning electron microscope image of the plasmonic nanoantennas (right).

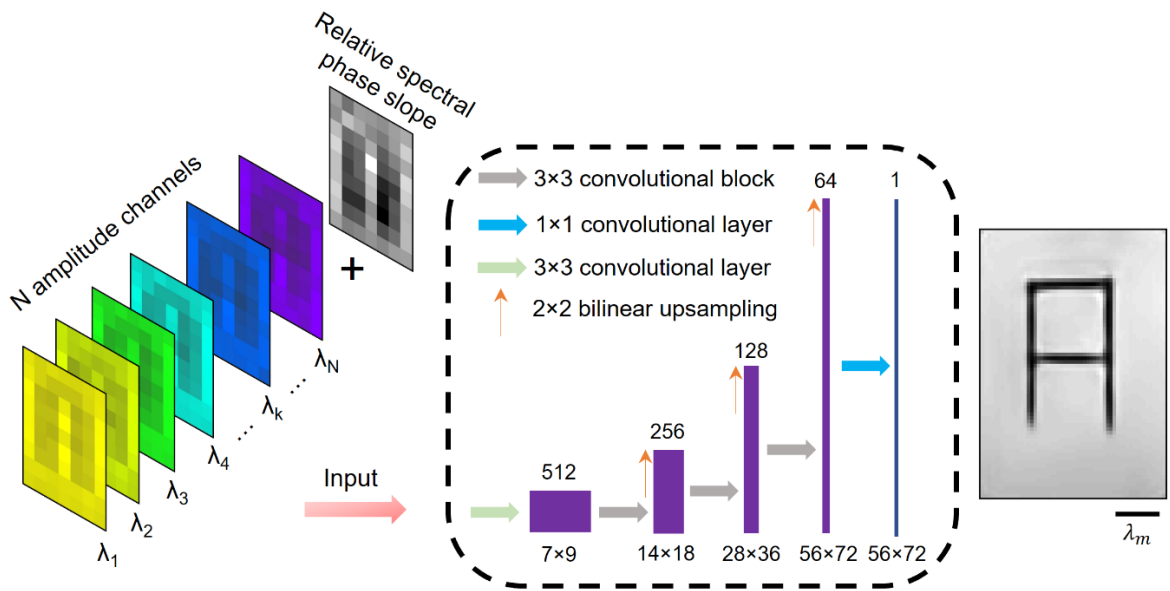

**a** Pixel super-resolution network

**b** Ground truth images

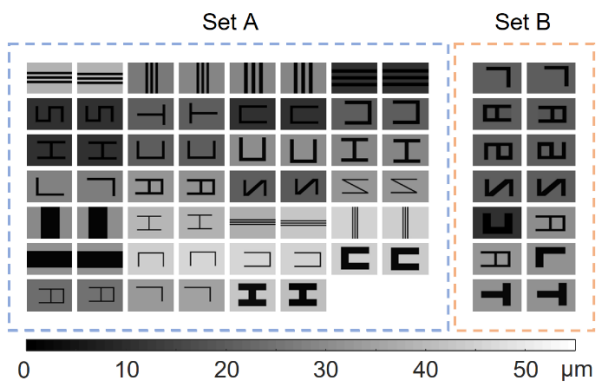

**c** Reconstructed super-resolved images

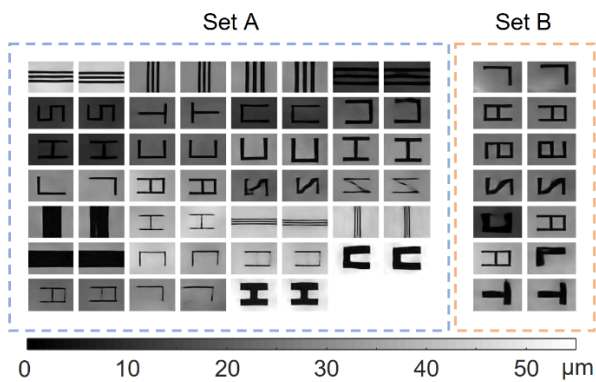

**d** Relative spectral phase slope

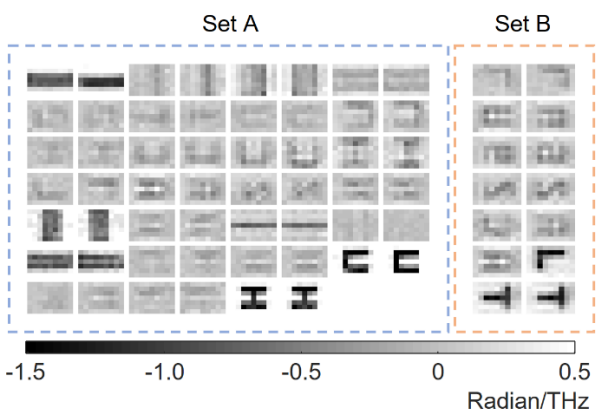

**e** Amplitude channel @ $\lambda_m$ (0.754 THz)

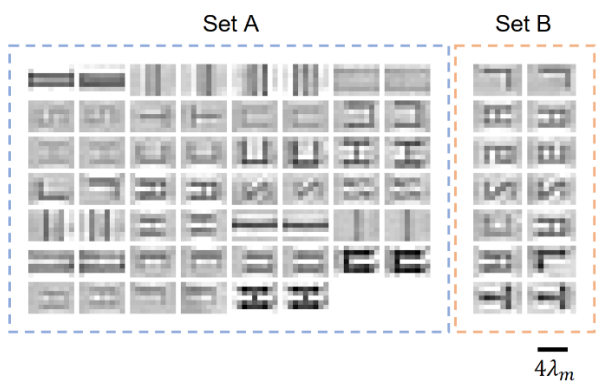

**Figure 2 PSR-enhanced THz-FPA imaging results. a,** The presented imaging system relies on the high SNR and broad detection bandwidth of our THz-FPA and exploits the spatiotemporal information collected in a single shot without any time-multiplexing or raster scanning, to enhance the resolution of the imaging system beyond the limit dictated by the physical size of the pixels with the help of a trained CNN. **b,** Ground truth thickness images of the objects used in our experiments. The blind testing Set A/B contains 54/14 measurements collected based on 27/8 different object patterns. Compared to the training set, the samples in Set A differ in terms of their thickness contrasts, whereas, the samples in Set B are unique regarding both their 2D patterns and thickness contrasts. **c,** Super-resolved images using the PSR-enhanced THz-FPA. The PSR neural network provides a 16-fold increase in the number of effective pixels, revealing >1 kilo pixels in each image. **d,** The relative spectral phase slope of the objects computed based on the slope of unwrapped spectral phase distribution. **e,** The amplitude channel at the median frequency, $f_m = 0.754$ THz.

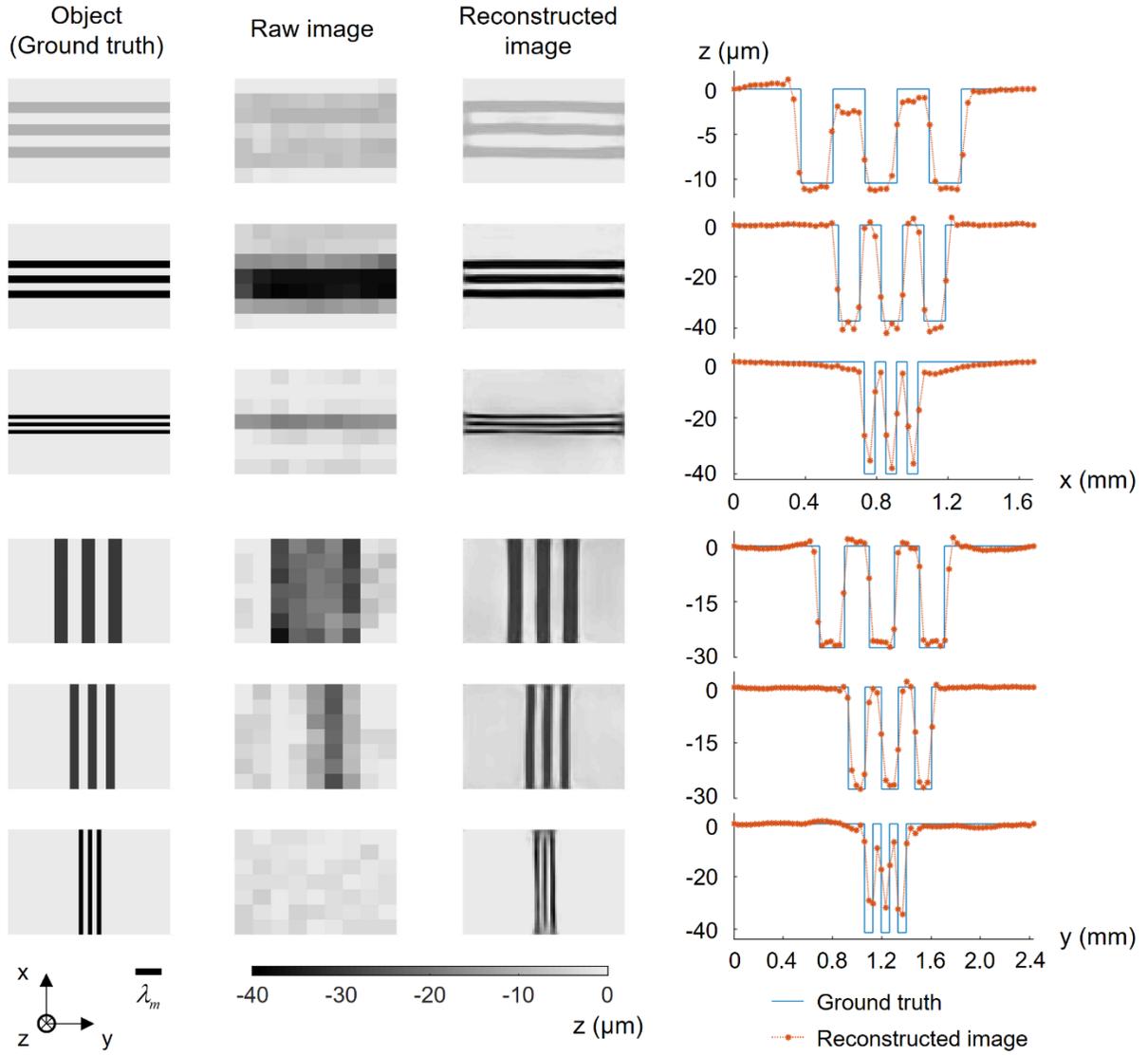

**Figure 3 Resolution quantification of the PSR-enhanced THz-FPA.** Resolved images of the fabricated resolution test objects are shown on the left. The linewidths and separations are 0.75×, 0.5×, and 0.25× of the FPA pixel size in both the horizontal and vertical directions. The raw images show the depth patterns estimated from the raw data of the THz-FPA. $\lambda_m$ = 398 μm denotes the median wavelength of the terahertz frequency band of interest. The right plots show the averaged depth profiles in the vertical (top 3 plots) and horizontal (bottom 3 plots) directions.

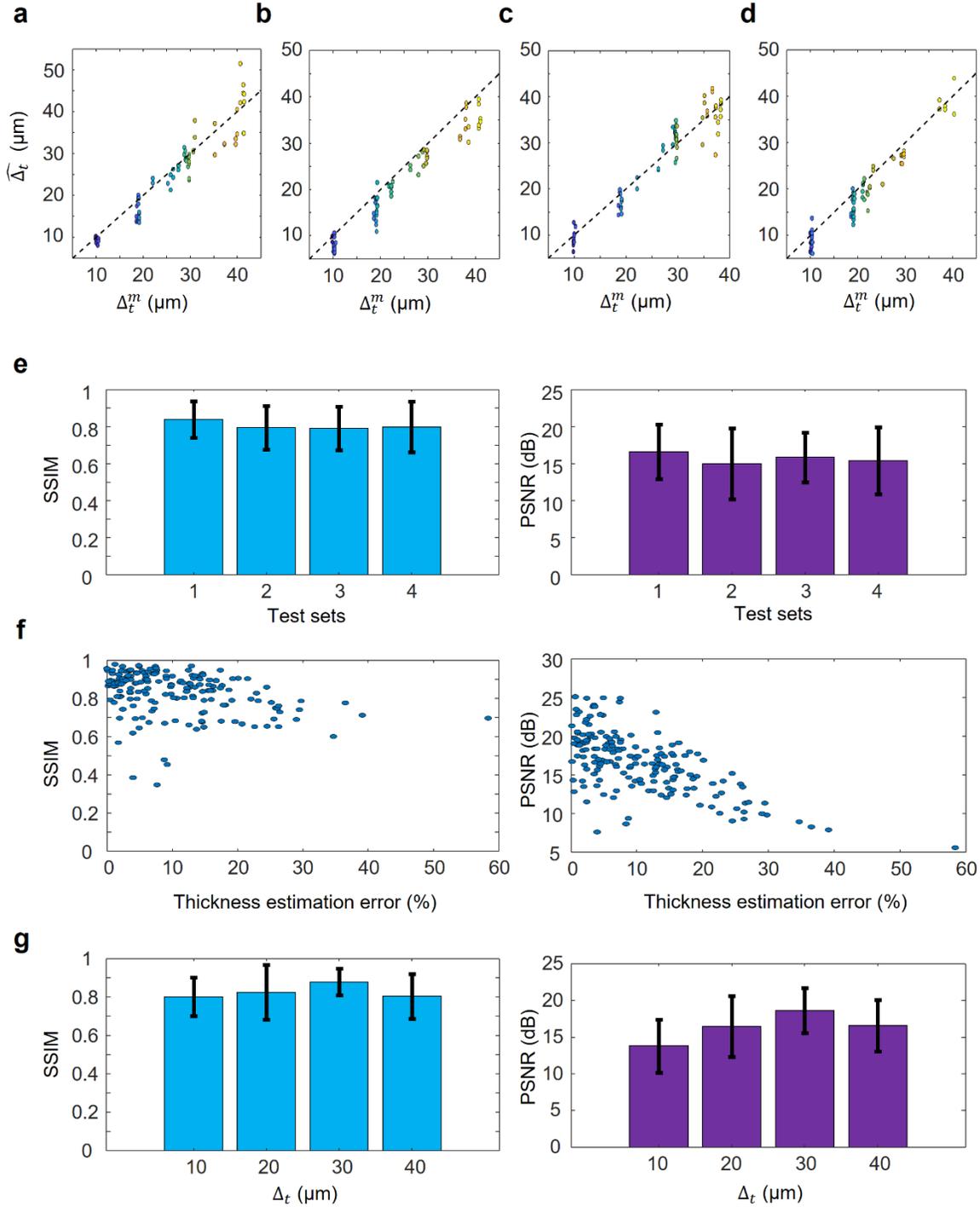

**Figure 4 Quantification of the image reconstruction quality and the thickness contrast prediction accuracy. a,** $\widehat{\Delta_t}$ vs. $\Delta_t^m$ for the test set shown in Fig. 2. The dashed black line represents $\widehat{\Delta_t} = \Delta_t^m$. **b-d,** Same as (a), except the reported $\widehat{\Delta_t}$ and $\Delta_t^m$ represent the estimated and true (ground truth) values of material thickness contrasts for the test image sets (b) Set 2A and Set B, (c) Set 3A and Set B, (d) Set 4A and Set B, which are illustrated in Supplementary Figs. S5a-c, respectively. **e,** The mean SSIM (left) and PSNR (right) values along with the corresponding standard deviations representing the image reconstruction quality achieved by the PSR networks. The only difference between these PSR networks is their partitioning of the training and testing sets (see Methods). **f,** Distribution of the SSIM (left) and PSNR (right) of all of the object patterns in the test sets Set A, Set B, Set 2A, Set 3A, and Set 4A as a function of the percentage of thickness estimation error, i.e., $\frac{|\widehat{\Delta}_t - \Delta_t^m|}{\Delta_t^m} \times 100$. **g,** The mean SSIM (left) and PSNR (right) values along with the corresponding standard deviations representing the image reconstruction quality achieved by the PSR networks for the object patterns included in the test sets Set A, Set B, Set 2A, Set 3A, and Set 4A as a function of $\Delta_t$.

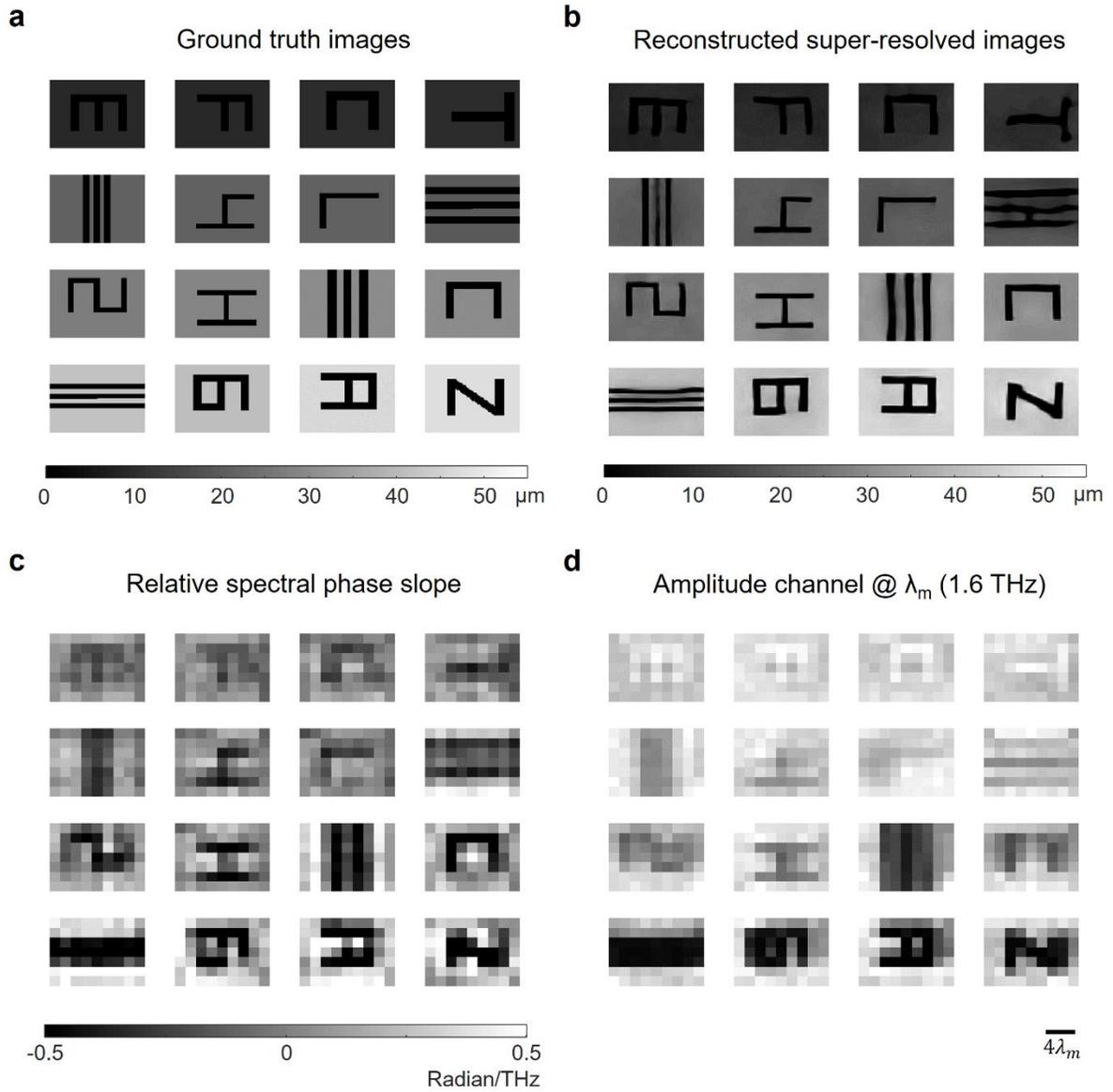

**Figure 5 Far-field PSR-enhanced THz-FPA imaging results. a,** Ground truth thickness images of the objects used in our experiments. **b,** Super-resolved images using the PSR-enhanced THz-FPA. **c,** The relative spectral phase slope of the objects computed based on the slope of unwrapped spectral phase distribution. **d,** The amplitude channel at the median frequency, $f_m$ = 1.6 THz.

**Methods**

1. THz-FPA fabrication process

The terahertz focal-plane array was fabricated on a low-temperature grown GaAs (LT-GaAs) substrate with a carrier lifetime of 0.3 ps (Fig. 1 inset). The fabrication process flow is shown in Supplementary Fig. S2. The plasmonic nanoantennas were patterned using electron-beam lithography, followed by the deposition of 3/47-nm Ti/Au and lift-off. The dipole-shaped nanoantennas have a 100 nm width and 180 nm periodicity along the z-axis and have an arm length and tip-to-tip gap size of 20 μm and 500 nm, respectively, along the x-axis. Output traces and contact pads were patterned by photolithography, followed by 20/180-nm Ti/Au deposition and lift-off. A 290-nm-thick silicon nitride anti-reflection coating was deposited by plasma-enhanced chemical vapor deposition. Shadow metals were formed on the anti-reflection coating by photolithography, 10/90-nm Ti/Au deposition, and lift-off. The fabricated THz-FPA consists of 283,500 plasmonic nanoantennas covering a 2.4 mm × 1.7 mm area.

2. Terahertz time-domain imaging setup

A Ti:sapphire laser (Coherent Mira 900) was used to generate optical pulses with an 800 nm central wavelength, 135 fs pulse width, and 76 MHz repetition rate. The optical beam was split into two branches. One branch pumped a plasmonic photoconductive terahertz emitter[46] to generate terahertz pulses. The other branch pumped the THz-FPA. The optical pump power incident on the terahertz emitter and FPA was 660 mW and 500 mW, respectively. For the experimental setup used for Fig. 2b-e, two parabolic mirrors were used to collimate and focus the generated terahertz radiation onto the active area of the THz-FPA after interacting with the imaged object. The objects were axially placed ~1.1 mm away from the active area of the THz-FPA using a high refractive index semiconductor substrate (see Supplementary Fig. S5a). For the experimental setup used for Fig. 5, the imaged objects were placed at a 20 cm axial distance from the active area of the THz-FPA. One parabolic mirror and one terahertz objective lens (TeraLens, Lytid) were used to collimate and focus the terahertz radiation onto the THz-FPA after interacting with the imaged objects while providing a demagnification factor of ~2.76 (see Supplementary Fig. S5b). An optical delay stage was used to vary the optical path difference between the optical pump and terahertz beams incident on the THz-FPA. The time-domain terahertz electric field was obtained by measuring the photocurrent of the FPA pixels as a function of the optical delay. The frequency-dependent amplitude and phase of the terahertz signal were calculated by taking the Fourier transform of the time-domain data.

3. THz-FPA data acquisition system

A custom-made readout circuit was built for data readout from the THz-FPA. The circuit consists of an FPGA development board (Basys 3, Digilent) controlling four 16-channel multiplexers (ADG1206, Analog Device), which direct the FPA outputs to a transimpedance amplifier (DLPCA-200, FEMTO) sequentially. The transimpedance amplifier converts the FPA photocurrent outputs to voltage signals, which are then sampled by a lock-in amplifier (MFLI, Zurich Instruments) and routed to a computer for post-processing. The THz-FPA outputs at each temporal point are captured in a 164 μs.

4. Phase object fabrication process

The phase objects were fabricated on a high-resistivity silicon substrate. The fabrication process flow is shown in Supplementary Fig. S7. A thin layer of silicon dioxide was first grown on the silicon substrate using thermal oxidation. Then the object patterns were transferred to the silicon dioxide layer by photolithography and silicon dioxide etching. Deep reactive-ion etching was used to create the trenches in the silicon substrate with the silicon dioxide layer being the etch mask. The depth of the silicon trenches was controlled by the etching time. Finally, the silicon dioxide layer was removed through buffered oxide etching. The depth of the silicon trenches was measured by Dektak 6M profilometer, which provides a 1 nm depth resolution.

5. Data preparation for deep neural network training and testing

The generation of ground truth images was automated using a MATLAB code that takes the etch mask (.dxf file) of the fabricated objects and outputs their thickness map with a resolution of 10 µm in both x and z directions, based on morphological image processing operations[47]. Since the relative position of the objects with respect to the active area of the THz-FPA is subject to change during the experiments, we implemented an algorithm that estimates the 2D relative location of the objects and accordingly shifts the high-resolution ground truth images. For the estimation of the object location, we followed a two-step procedure. In the first step, the high-resolution ground truth images were down-sampled to a Cartesian grid defined by the physical pixel size of our FPA, i.e., 240 µm × 270 µm. For each spectral component detected within the 0.5-1 THz range, we computed the normalized cross-correlation function between the detected amplitude channels and the down-sampled, low-resolution ground truth model. The spectral component exhibiting the highest correlation coefficient was selected as the reference for the second fine adjustment step, and the location of the peak correlation coefficient was used for the coarse alignment.

Since an image alignment accuracy beyond the physical pixel size was required to train the PSR networks accurately, we adapted a greedy strategy for fine alignment. Each ground truth high-resolution image was shifted to 24 × 27 different positions with 10 µm steps around its coarsely estimated location computed in the first step. At each position, the ground truth was down-sampled using bilinear interpolation to the physical resolution of the FPA and compared to the amplitude channel of the reference spectral component determined in the first step. Out of 24 × 27 = 648 normalized cross-correlations, we determined the best alignment position as the one providing the highest correlation coefficient, providing us an estimate of the relative object location with respect to the FPA at a spatial sampling rate much higher than the size of the physical pixels. All of these steps constitute a one-time effort used for the training of our PSR networks.

To super-resolve the images of the spatially structured silicon samples placed at an axial distance of 1.1 mm from the THz-FPA, we constructed a dataset containing 343 measurements, and out of these 343 measurements, we reserved 68 measurements corresponding to 35 unique samples for blind testing (see Fig. 2b), while the remaining 275 measurements were used to train the PSR network. In these 68 measurements constituting the blind testing dataset shown in Fig. 2, there were two major groups of objects labeled as Set A, 54 measurements, and Set B, 14 measurements. The 54 measurements contained in Set A were distributed over 27 unique object patterns, and for each of these object patterns, the training set contained measurements produced by the same 2D pattern albeit fabricated with different etch thickness levels. The samples in Set B, on the other hand, did not only differ from the training samples in terms of the etch thickness contrast, but their 2D patterns were also unique, not included in the training set. Preserving this training/testing partition ratio, 275/68, we also created 3 additional datasets shown in Supplementary Fig. S9. While the samples in Set B of these 3 new blind testing datasets were the same as the one shown in Fig. 2, the samples in Set A were different and complementary to each other in terms of material thickness contrasts. To super-resolve the images of the spatially structured silicon samples placed at an axial distance of 20 cm from the plane of the THz-FPA, we constructed an experimental dataset containing 151 far-field measurements, reserved 16 measurements corresponding to 16 unique samples for blind testing, and used the remaining 135 measurements to train the PSR neural network.

6. PSR network architecture

Figure 2a illustrates the architecture of the deep neural network for PSR. In this study, we used a CNN inspired by the up-sampling arm of U-Net[48]. Although we investigated the performance of other CNN architectures proven to provide high-quality super-resolved images, e.g., SRResNet[49], SRCNN[50], and ESPCN[51], the architecture shown in Fig. 2a outperformed these architectures based on our THz-FPA results.

Each gray arrow in Fig. 2a represents a convolutional block that consists of 3 convolutional layers. If we denote the number of channels at the input (output) of a convolutional block, $c$, by $N_c^i$ ($N_c^o$), then the number of channels is decreased at the output by half, i.e., $N_c^o = \frac{N_c^i}{2}$. This reduction in the number of channels is achieved by the 3 convolutional layers in $c^{th}$ block, where the number of channels at the (input, output) of each of these convolution operations is given as, $\left(N_c^i, \frac{N_c^i+N_c^o}{2}\right)$, $\left(\frac{N_c^i+N_c^o}{2}, \frac{N_c^i+N_c^o}{2}\right)$ and $\left(\frac{N_c^i+N_c^o}{2}, N_c^o\right)$, respectively. In addition, between each convolutional block, there is a 2 × 2, bilinear up-sampling operation. We used 4 convolutional blocks interconnected through 3 linear upsampling operations. Including the first convolutional layer that takes $N_\lambda$ amplitude channels concatenated with the 7 × 9 low-resolution thickness estimate, $I_t^{LR}$, and the output layer (blue arrow in Fig. 2a), our neural network consists of, in total, 14 convolutional layers.

To overcome overfitting, we introduced data augmentation as part of the training forward model, in the form of random vertical and horizontal flipping of the input amplitude and thickness estimate channels. In addition, we employed random 90° rotations, with an additional re-sampling procedure (using bilinear interpolation) due to the nonsymmetric structure of our FPA considering the number of pixels in x and z directions. While the random flipping and 90° rotations prevented the overfitting and improved the generalization performance of the PSR network, to further enhance our super-resolution imaging performance, we adapted a simple yet effective data augmentation technique, i.e., MixUp[52,53]. The MixUp technique creates linear combinations of input and output pairs based on a random variable, $\delta$, in the range [0,1] drawn from $Beta(\alpha, \alpha)$ distribution. In this study, $\alpha$ was taken as 0.2. With $X_q$ and $X_p$ denoting the input data sensors for the $q^{th}$ and $p^{th}$ experimental measurements, the MixUp technique created a new input, $X_{qp}$,

$$X_{qp} = \delta X_q + (1-\delta)X_p. \qquad 1$$

and similarly, generated/augmented the corresponding ground truth image, $I_{qp}^{SR}$,

$$I_{qp}^{SR} = \delta I_q^{SR} + (1-\delta)I_p^{SR}. \qquad 2$$

where, $I_q^{SR}$ and $I_p^{SR}$ represent the ground truth images corresponding to the data tensors $X_q$ and $X_p$, respectively. All the ground truth images, $I^{SR}$, are measured by the Dektak 6M profilometer.

We trained our PSR networks using a structural loss function, $L$, based on the mean-squared-error (MSE),

$$L = \frac{1}{MN}\sum_{m=1}^{M}\sum_{n=1}^{N}\left|I^{SR}[m,n] - I^{SR'}[m,n]\right|^2. \qquad 3$$

where, $I^{SR'}$ defines the inference of the PSR network, with $M \times N$ output pixels.

The image quality metric, PSNR, which was directly related to the MSE, was computed based on,

$$PSNR = 20\log_{10}\left(\frac{T}{\sqrt{L}}\right), \qquad 4$$

where, $T$ denotes the etch depth of the ground truth object measured by the Dektak 6M profilometer.

To compute the other image quality metric, SSIM, we used the following equation,

$$SSIM = \frac{1}{W}\sum_{w=1}^{W}\frac{\left(2\mu_{I_w^{SR}}\mu_{I_w^{SR'}} + C_1\right)\left(2\sigma_{I_w^{SR}I_w^{SR'}} + C_2\right)}{\left(\mu_{I_w^{SR}}^2 + \mu_{I_w^{SR'}}^2 + C_1\right)\left(\sigma_{I_w^{SR}}^2 + \sigma_{I_w^{SR'}}^2 + C_2\right)}, \qquad 5$$

where the terms $\mu_{I_w^{SR}}$ and $\mu_{I_w^{SR'}}$, denote the local mean computed for each 8×8 window, $w$, scanned over the ground truth thickness image, $I^{SR}$ and the super-resolved image at the output of the PSR network, $I^{SR'}$, respectively. Similarly, $\sigma_{I_w^{SR}}$ and $\sigma_{I_w^{SR'}}$ denote the corresponding local standard deviations with $\sigma_{I_w^{SR}I_w^{SR'}}$ representing the local cross-covariance between corresponding windows from $I^{SR}$ and $I^{SR'}$. As suggested in its original implementation[54], we convolved both $I^{SR}$ and $I^{SR'}$ with an 11×11 truncated, circularly symmetric Gaussian filter with a standard deviation of 1.5 before computing Eq. 5. The constant factors in Eq. 5, i.e., $C_1$ and $C_2$, were computed as $(K_1T)^2$ and $(K_2T)^2$, with $T$ denoting the ground truth material thickness contrast. The multiplicative terms $K_1$ and $K_2$ were set to be 0.01 and 0.03, respectively following the original implementation[54].

**Methods References**

53. Zhang, H., Cisse, M., Dauphin, Y. N. & Lopez-Paz, D. mixup: Beyond Empirical Risk Minimization. *arXiv:1710.09412 [cs, stat]* (2018).

54. Wang, Z., Bovik, A. C., Sheikh, H. R. & Simoncelli, E. P. Image quality assessment: from error visibility to structural similarity. *IEEE Transactions on Image Processing* **13**, 600–612 (2004).

# Supplementary materials for

# Plasmonic photoconductive terahertz focal-plane array with pixel super-resolution

**Authors:** Xurong Li[1,2] †, Deniz Mengu[1,2,3] †, Aydogan Ozcan[1,2,3] and Mona Jarrahi[1,2,*]

**Affiliations:**

[1]Department of Electrical & Computer Engineering, University of California Los Angeles (UCLA), California, USA

[2]California NanoSystems Institute (CNSI), University of California Los Angeles (UCLA), California, USA

[3]Department of Bioengineering, University of California Los Angeles (UCLA), California, USA

*Corresponding author. Email: mjarrahi@ucla.edu

† Equal contribution

**Supplementary Text:**

**Impact of the number of nanoantenna clusters on the performance of the FPA**

For a given number of plasmonic nanoantennas, the number of clusters, $N_{clusters}$, determines both the SNR and space-bandwidth product (and also the spatial resolution within a given field-of-view) of the imaging system. The SNR of each cluster is calculated as $I_{ph}^2/\overline{i_n^2}$, where $I_{ph}$ is the induced photocurrent in each cluster in response to an incident terahertz radiation and $i_n$ is the noise current of each cluster, dominated by the Johnson-Nyquist noise. The induced photocurrent and noise current are proportional to $P_{opt}$ and $P_{opt}^{1/2}$, respectively, where $P_{opt}$ is the optical power pumping each cluster. For a uniform optical illumination across all the FPA nanoantennas, the optical power pumping each cluster is calculated as the total optical pump power divided by the number of clusters. Therefore, the SNR of each cluster is inversely proportional to the number of clusters, $SNR \propto 1/N_{clusters}$.

The number of clusters also determines the native space-bandwidth product of the FPA (unless a computational reconstruction method is used to further enhance it as we demonstrated in this work). Stated differently, the spatial resolution of the native terahertz imaging system within a given sample field-of-view is inversely proportional to (and gets better with) the number of clusters in each lateral direction. For the fabricated FPA with ~0.3 million plasmonic nanoantennas, the number of clusters was chosen to be 63 to provide a very high SNR for all the clusters (81 dB peak SNR) while maintaining an effective pixel number of 1-kilo pixels after the neural network-based PSR. Assuming that the total number plasmonic nanoantennas on the FPA is fixed/constant (e.g., 283,500 as used in this work), further increasing the number of clusters would improve the native space-bandwidth product of the FPA at the cost of an SNR degradation. However, a larger number of clusters can be achieved at the same SNR level (81 dB peak SNR) by further increasing the number of the plasmonic nanoantennas to e.g., >1 million, which is within the reach of the current fabrication methods to further scale-up the presented FPA technology in terms of its effective space-bandwidth product.

**Impact of spectral bandwidth on the performance of the PSR network**

The super-resolved thickness images shown in Fig. 2 of the main text and Supplementary Fig. S5 were computed by the associated convolutional neural networks (CNNs) that used the 41 amplitude channels of the spectral components within the band 0.5 THz – 1 THz. In addition, the $42^{nd}$ channels at the input of these PSR networks representing the estimate of the thickness images were also computed based on the slope of the unwrapped phase within the same band. To select the optimal band for estimating the slope of the spectral phase and the amplitude channels fed into the convolutional PSR network architecture shown in Fig. 2, we relied on empirical evidence shown in Supplementary Figs. S9 and S10, respectively. In Supplementary Fig. S9, we compared the impact of the selected band of frequencies on the quality of the thickness image estimate computed based on the slope of the spectral phase in terms of squared-error (MSE) and SSIM computed with respect to the ground truth images. Although the behavior of SSIM and MSE exhibits a slightly different behavior as a function of bandwidth, a common trend is that slope-based material thickness estimation yields better results for both metrics as the spectral bandwidth utilized for fitting a line to the unwrapped phase increases. The MSE metric, on the other hand, shows a critical turning point for the band between 0.5 THz – 1.1 THz, where, contrary to the general trend, the inclusion of the components from 1 THz to 1.1 THz causes an increase in the error due to the water absorption lines located in the close vicinity of 1.1 THz. Therefore, we opted to use the spectral band between 0.5 THz and 1 THz to estimate the slope of the spectral phase which, in return, results in the estimated thickness images fed into the CNNs.

Next, we investigated the effect of the bandwidth covered by the amplitude channels on the PSR performance of our terahertz imaging system. Towards this goal, we swept the spectral bandwidth contributing to the input channels of the PSR neural network for the test set shown in Fig. 2 across 8 different candidates starting with the narrowest-band option 0.5 THz – 0.55 THz up to the widest using all the components within the range of 0.5 THz – 2.1 THz. For each of these candidate bandwidths, we trained a new PSR network from scratch and compared the quality of the super-resolved images in terms of both SSIM and PSNR as shown in Supplementary Fig. S10. Similar to the trend in Supplementary Fig. S9, both SSIM and PSNR improve as the bandwidth, thus the number of amplitude channels at the input of the convolutional network, increases up to the point, where all the spectral components between 0.5 THz and 1 THz are utilized. For instance, the deep neural network trained based on the smallest bandwidth considered in this analysis covering the range from 0.5 THz to 0.55 THz can only achieve 0.737±0.137 SSIM score and 13.08±5.09 dB PSNR value. These values point to a significantly inferior image reconstruction performance compared to the images shown in Fig. 2 for which the SSIM and PSNR were reported in the main text as 0.839±0.098 and 16.60±3.67 dB, respectively, highlighting the vital role of the broadband operation of our THz-FPA in achieving the super-resolution imaging task.

**Numerical forward model of the lensless terahertz imaging setup**

The terahertz numerical forward model represents the terahertz wave propagation and its interaction with the imaged objects, and generates the hyperspectral output images at the FPA plane corresponding to arbitrary objects. We formulated the terahertz forward model using the Rayleigh-Sommerfeld diffraction integral and the related angular spectrum representation of terahertz waves. Let $u$ and $w$ column vectors (N×1) represent the sampled terahertz fields (including the phase and amplitude information) at the input and output of a homogeneous medium, respectively. These two vectors are related through a matrix multiplication, $H_{d,n,\lambda} u = w$, where $H_{d,n,\lambda}$ is an N×N matrix that represents the Rayleigh-Sommerfeld diffraction between two fields specified over parallel planes, which are axially separated by a distance $d$. The subscripts $n$ and $\lambda$ denote the refractive index of the propagation medium and the wavelength of the diffracted terahertz wave, respectively. Since a homogeneous medium acts as a linear shift-invariant system, $H_{d,n,\lambda}$ is a Toeplitz matrix, and is diagonalizable by the 2D Fourier transform matrix, $F$, i.e., $H_{d,n,\lambda} = F^{-1} Q_{d,n,\lambda} F$, and its eigenvalues (i.e., the diagonal elements of $Q_{d,n,\lambda}$), $q(\upsilon, \nu, d, n, \lambda)$, can be written as,

$$q(\upsilon, \nu, d, n, \lambda) = \left\{ \begin{array}{ll} exp\left[j 2\pi \frac{nd}{\lambda} \sqrt{1 - \left(\frac{\lambda}{n}\upsilon\right)^2 - \left(\frac{\lambda}{n}\nu\right)^2}\right] & for\ \upsilon^2 + \nu^2 \leq \left(\frac{n}{\lambda}\right)^2 \\ 0 & otherwise \end{array} \right\}. \quad (S1)$$

where, $\upsilon$ and $\nu$ denote the spatial frequencies.

The fabricated thickness patterns used in our experiments were etched over the surface of a 500-μm-thick silicon substrate and the volume between the back-surface of the silicon substrate and the active area of the THz-FPA is covered by a 680-μm-thick GaAs substrate. Therefore, the wave propagation in our terahertz forward model uses matrices $Q_{d_{GaAs}, n_{GaAs}, \lambda}$ and $Q_{d_{Si}, n_{Si}, \lambda}$ containing the eigenvalues of wave diffraction inside these two materials. Accordingly, the terahertz transfer function from the etched surface of the silicon substrate (i.e., the object plane) to the active area of the THz-FPA is given by $T_\lambda = F^{-1} Q_{d_{GaAs}, n_{GaAs}, \lambda} Q_{d_{Si}, n_{Si}, \lambda} F$. Since the maximum etch thickness, $\Delta_t$, used in our experimental system is 40 μm that is much smaller than the wavelengths considered in our forward model (i.e., $\Delta_t \ll \lambda_k, for\ k = 1,2,3, \dots, N$), we modeled the object pattern, $t[p, r]$, as a thin modulation surface. Consequently, for the sampled illumination beam at a given wavelength, $i_\lambda[p, r]$, incident over the etched surface of the silicon substrate with the etching pattern described by $t[p, r]$, the outgoing, modulated beam is given by $o_\lambda[p, r] = i_\lambda[p, r] \exp(-j 2\pi \frac{(n_{Si}-1)t[p,r]}{\lambda})$, where $n_{Si}$ is the refractive index of silicon. $n_{Si}$ and $n_{GaAs}$ were assumed to be 3.42 and 3.6, respectively.

In our forward model, we represented the input thickness patterns and the associated terahertz fields at every plane with a sampling rate of 30 μm and 33.75 μm in the lateral dimensions, which correspond to 1/8[th] of the physical pixel size of our THz-FPA, 240μm×270μm. Stated differently, a given thickness pattern $t[p,r]$ is represented through 56×72 pixels and each pixel was assumed to occupy an area of 30 μm×33.75 μm.

According to our forward model, for a thickness pattern $t[p,r]$ of size 56×72 at any given wavelength, the corresponding complex-valued field measured by our THz-FPA can be written as $\boldsymbol{u}_\lambda[s'] = \boldsymbol{D}\boldsymbol{T}_\lambda' \boldsymbol{o}_\lambda[s]$, where $\boldsymbol{o}_\lambda[s]$ denotes the 1D vectorized counterpart of $\boldsymbol{o}_\lambda[p,r]$, i.e., $\boldsymbol{o}_\lambda[s] = vec(\boldsymbol{o}_\lambda[p,r])$. $\boldsymbol{T}_\lambda'$ is a (56×72)×(56×72) matrix obtained from the N×N terahertz transfer matrix $\boldsymbol{T}_\lambda$ described above by deleting the appropriately selected rows and columns. In our simulations N was taken as 512. $\boldsymbol{D}$ is a matrix of size (7×9)×(56×72) representing: (1) the integration of the complex-valued field samples over each pixel of the THz-FPA, and (2) the downsampling operator.

In order to expand the deep learning-based super-resolved material thickness estimation scheme presented in our paper and cover object patterns with multiple levels of etch depths and smaller contrast, we simulated the wave fields at 41 wavelengths from 0.5 THz to 1 THz created by 1092 different object patterns with non-binary depths and smaller contrast, using the above outlined numerical forward model. Out of these 1092 hyperspectral simulated object fields, we used 984 for transfer learning that further trained the neural network shown in Fig. 2a of the main text (originally trained only on the experimental data) to cover these new set of objects with multiple etch depths. During the transfer learning phase, the learning rate was set to be 200× smaller and the number of epochs was set to be 2.5× larger compared to the original training based solely on the experimental data. The remaining 108 simulated hyperspectral field data were reserved for blind testing to quantify the success of the transfer learning step and the resulting neural network model. Comparison of the predictions of our terahertz forward model against the experimental measurements are shown in Supplementary Fig. S17, demonstrating the agreement between the generated/simulated output images at the FPA plane and the corresponding experimental results.

**Holographic reconstruction of the terahertz video frames**

To reconstruct each frame of the terahertz video, we take advantage of the coherent wave detection capabilities of our THz-FPA. Specifically, since our THz-FPA can measure both the amplitude and phase of terahertz radiation at each pixel, we can holographically reconstruct the input objects. Hence, the problem of reconstructing an input object/scene reduces to inverting the terahertz wave transfer matrix between the complex-valued measurements, $\boldsymbol{u}_\lambda$, and the complex-valued object modulation function, $\boldsymbol{o}_\lambda$. Denoting the targeted super-resolution factor in our multi-wavelength holographic reconstruction procedure with $\alpha_{SR}$, the wave transfer matrix, $\boldsymbol{A}_\lambda$, is of size (7×9)×(7$\alpha_{SR}$×9$\alpha_{SR}$), at a given wavelength, $\lambda$. $\boldsymbol{A}_\lambda$ can be described as $\boldsymbol{A}_\lambda = \boldsymbol{D}\boldsymbol{T}_\lambda'$, where $\boldsymbol{D}$ is a matrix of size (7×9)×(7$\alpha_{SR}$×9$\alpha_{SR}$) representing: (1) the integration of complex-valued field samples over each pixel of our THz-FPA, and (2) the downsampling operator. The matrix $\boldsymbol{T}_\lambda'$, on the other hand, is the (7$\alpha_{SR}$×9$\alpha_{SR}$)×(7$\alpha_{SR}$×9$\alpha_{SR}$) inner matrix of an N×N wave propagation matrix, $\boldsymbol{T}_\lambda$, which is defined as $\boldsymbol{T}_\lambda = \boldsymbol{F}^{-1} \boldsymbol{Q}_{d_{GaAs}, n_{GaAs}, \lambda} \boldsymbol{Q}_{d_{tape}, n_{tape}, \lambda} \boldsymbol{F}$. The diagonal matrices $\boldsymbol{Q}_{d_{GaAs}, n_{GaAs}, \lambda}$ and $\boldsymbol{Q}_{d_{tape}, n_{tape}, \lambda}$ contain the eigenvalues of wave propagation inside the 680-μm-thick GaAs substrate and 75-μm-thick sticky tape used in our setup. For the reconstructed video shown in Supplementary Video and Supplementary Fig. S21, the super-resolution factor, $\alpha_{SR}$, was taken as 2 corresponding to 2×2, i.e., 4× increase in the space-bandwidth product of our THz-FPA. For any super-resolution factor greater than 1, $\alpha_{SR}>1$, the $A_\lambda$ is a fat matrix, in other words, it has more columns than rows. As a result, $A_\lambda$ is not invertible. To be able to solve this holographic super-resolved image reconstruction problem, we constructed a larger matrix $A = \begin{bmatrix} A_{\lambda_1} \\ A_{\lambda_2} \\ \cdots \\ A_{\lambda_8} \end{bmatrix}$ of size $(8\times7\times9)\times(7\alpha_{SR}\times9\alpha_{SR})$ that contains the transfer matrices of 8 different spectral components from 0.4 THz to 1 THz. Similarly, a measurement vector, $u$, was constructed as follows; $u = \begin{bmatrix} u_{\lambda_1} \\ u_{\lambda_2} \\ \cdots \\ u_{\lambda_8} \end{bmatrix}$ containing the complex-valued field measured by our THz-FPA for all these 8 wavelength components. The forward model of the system can be written as $Ao = u$, where $o$ is the $(7\alpha_{SR}\times9\alpha_{SR})\times1$ vector containing the complex-valued object field. $A$ contains more rows than columns which suggests that the solution that minimizes the function $f(o) = \|Ao - u\|^2$ can be solved using the pseudo-inverse of $A$. However, despite being a tall matrix, $A$ is an ill-conditioned matrix which inevitably amplifies the noise component in the measurement vector $u$. Consequently, to reliably compute the solution that minimizes $f(o)$, we regularized the solution space. In this work, we achieved this by solving

$$f'(o) = \|Ao - u\|^2 + \rho\|o\|_{TV} + \xi\|g(\angle o)\|_{TV} \qquad (S2)$$

where $\|.\|_{TV}$ stands for the total-variation (TV) norm. In our algorithm, we used the isotropic definition of TV norm, $\|x\|_{TV} = \sum_p \sum_r \sqrt{(\Delta_p^h x)^2 + (\Delta_p^v x)^2}$, with $\Delta_p^h$ and $\Delta_p^v$ denoting the horizontal and vertical (on the 2-D lattice) first-order local difference operators (omitting the boundary corrections) [55]. The function $g$ represents a 2D phase unwrapping operator, which, in this work, was implemented based on [56]. For each frame of the video illustrating the water flow through the pipes, the solution of this equation was computed using the Two-step Iterative Shrinkage-Thresholding (TwIST) algorithm [57] with $\rho = 10^{-4}$ and $\xi = 3 \times 10^{-5}$, which determine the strength of the TV-norm regulation on the amplitude and phase-channels of the complex-valued amplitude of the object field, $o$, respectively (see Eq. S2).

| Ref. | Year | Object imaging | Laser amplifier | Laser optical pulse energy | Pulse energy for THz detection | Pulse energy for THz generation | Laser repetition rate | Peak SNR |
|---|---|---|---|---|---|---|---|---|
| [58] | 1996 | No | NS | NS | NS | NS | NS | NS |
| [59] | 1997 | Yes | Yes | 4 µJ | NS | 2 µJ | 250 kHz | NS |
| [60] | 1998 | Yes | Yes | NS | NS | NS | NS | 40 dB |
| [61] | 2000 | Yes | Yes | 4 µJ | NS | NS | 250 kHz | NS |
| [62] | 2002 | Yes | Yes | > 800 µJ | 1.4 µJ | 800 µJ | 1 kHz | NS |
| [63] | 2003 | No | Yes | > 150 µJ | NS | 150 µJ | 1 kHz | NS |
| [64] | 2004 | No | Yes | > 140 µJ | NS | 140 µJ | 1 kHz | 56.7 dB |
| [65] | 2004 | No | Yes | > 260 µJ | 60 µJ | 200 µJ | 1 kHz | 46.8 dB |
| [66] | 2004 | No | Yes | NS | NS | NS | 1 kHz | NS |
| [67] | 2005 | Yes | Yes | > 575 µJ | 75 µJ | 500 µJ | 1 kHz | 34.0 dB |
| [68] | 2005 | Yes | Yes | NS | NS | NS | 1 kHz | NS |
| [69] | 2006 | Yes | Yes | 700 µJ | NS | NS | 1 kHz | NS |
| [70] | 2007 | Yes | Yes | NS | NS | NS | 1 kHz | NS |
| [71] | 2009 | Yes | No | 25 nJ | 3.4 nJ | 17.1 nJ | 76 MHz | 40 dB |
| [72] | 2009 | Yes | Yes | 900 µJ | 10 µJ | 890 µJ | 1 kHz | 46.8 dB |
| [73] | 2009 | Yes | Yes | 650 µJ | NS | NS | 1 kHz | 46.0 dB |
| [74] | 2010 | Yes | Yes | > 510 µJ | 10 µJ | 500 µJ | 1 kHz | NS |
| [75] | 2010 | Yes | Yes | 800 µJ | NS | NS | 1 kHz | NS |
| [76] | 2010 | Yes | Yes | 800 µJ | NS | NS | 1 kHz | 45.1 dB |
| [77] | 2011 | Yes | No | 25 nJ | NS | NS | 76 MHz | NS |
| [78] | 2011 | Yes | Yes | 4 mJ | NS | NS | 1 kHz | 40.0 dB |
| [79] | 2011 | Yes | Yes | 4 mJ | NS | 3.5 mJ | 1 kHz | 30.0 dB |
| [80] | 2012 | Yes | Yes | 4 mJ | NS | NS | 1 kHz | NS |
| [81] | 2013 | Yes | Yes | 800 µJ | 10 µJ | 490 µJ | 1 kHz | NS |
| [82] | 2013 | Yes | Yes | NS | 350 µJ | NS | NS | NS |
| [83] | 2016 | Yes | Yes | 4 mJ | NS | NS | 1 kHz | 48.5 dB |
| [84] | 2019 | Yes | Yes | 900 µJ | 20 µJ | 880 µJ | 1 kHz | NS |
| [85] | 2022 | Yes | Yes | 4 mJ | NS | NS | 1 kHz | NS |
| **Our Work** | 2022 | Yes | No | 31.6 nJ | 8.7 nJ | 6.6 nJ | 76 MHz | 81.4 dB |

**Table S1.** Specifications of the existing terahertz imaging systems without raster scanning through electro-optic detection and their comparison with the specifications of our imaging system (last row) based on a plasmonic photoconductive THz-FPA. The specifications listed as 'NS' are Not Specified in the listed references. The references that did not demonstrate object imaging, instead imaged the terahertz beam profile.

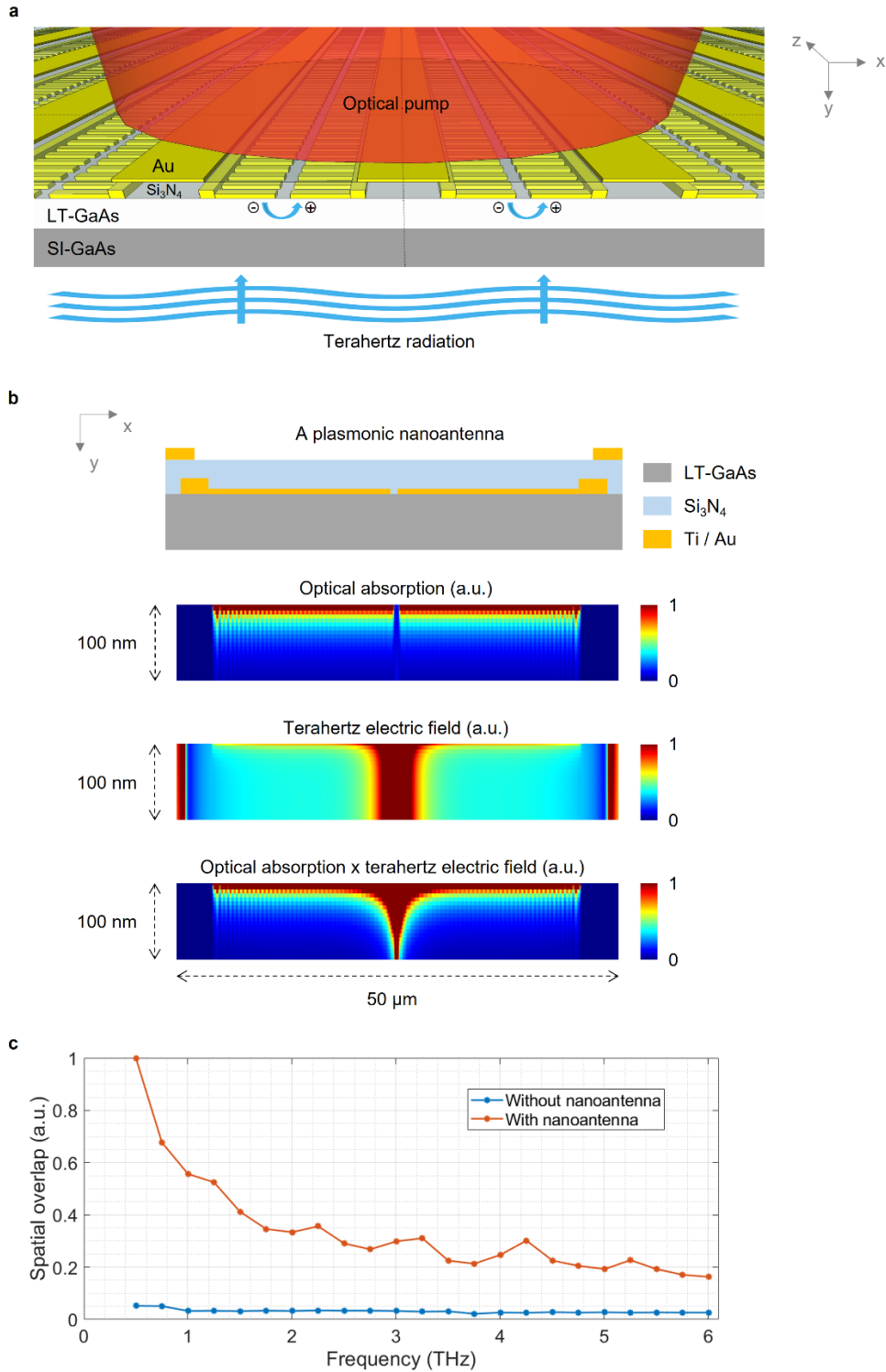

**Figure S1 a,** Schematic diagram of the THz-FPA based on a distributed plasmonic nanaoantenna array architecture. The geometry of the nanoantennas is chosen to obtain a strong spatial overlap between the received terahertz radiation and optical pump beam to achieve high detection sensitivity over a broad terahertz frequency range. **b,** The optical absorption profile (at $\lambda = 800$ nm), terahertz electric field profile (at $f = 1$ THz), and the multiplication product of the optical absorption and terahertz electric field profiles within a 100 nm depth in the LT-GaAs substrate. **c,** Spatial overlap between the optical and terahertz beams as a function of terahertz frequency with and without the plasmonic nanoantennas, which is defined as the integral of the multiplication product of the optical absorption and terahertz electric field under the nanoantennas. The distributed plasmonic nanaoantenna array architecture for the THz-FPA provides a much higher optical fill factor (49.1% for the demonstrated THz-FPA) compared to conventional photoconductive terahertz antenna arrays (less than 1%). In conventional photoconductive terahertz antennas, small optical active areas are connected to much larger discrete terahertz antenna elements (e.g., dipole, bow-tie, and spiral antennas), leading to a low optical fill factor when used in arrays. Since the detected terahertz power has a quadratic dependence on the optical pump power incident on the device active area, which is proportional to the optical fill factor, our THz-FPA provides significantly higher SNR (81 dB peak SNR) and broader bandwidth compared to conventional photoconductive terahertz antenna arrays with discrete terahertz antenna elements [86, 87]. To provide a high optical fill factor, the THz-FPA does not require any special focusing optics (e.g., microlens arrays, fiber bundles, or diffractive elements), making this distributed plasmonic nanoantenna array architecture scalable for large-pixel-count THz-FPAs.

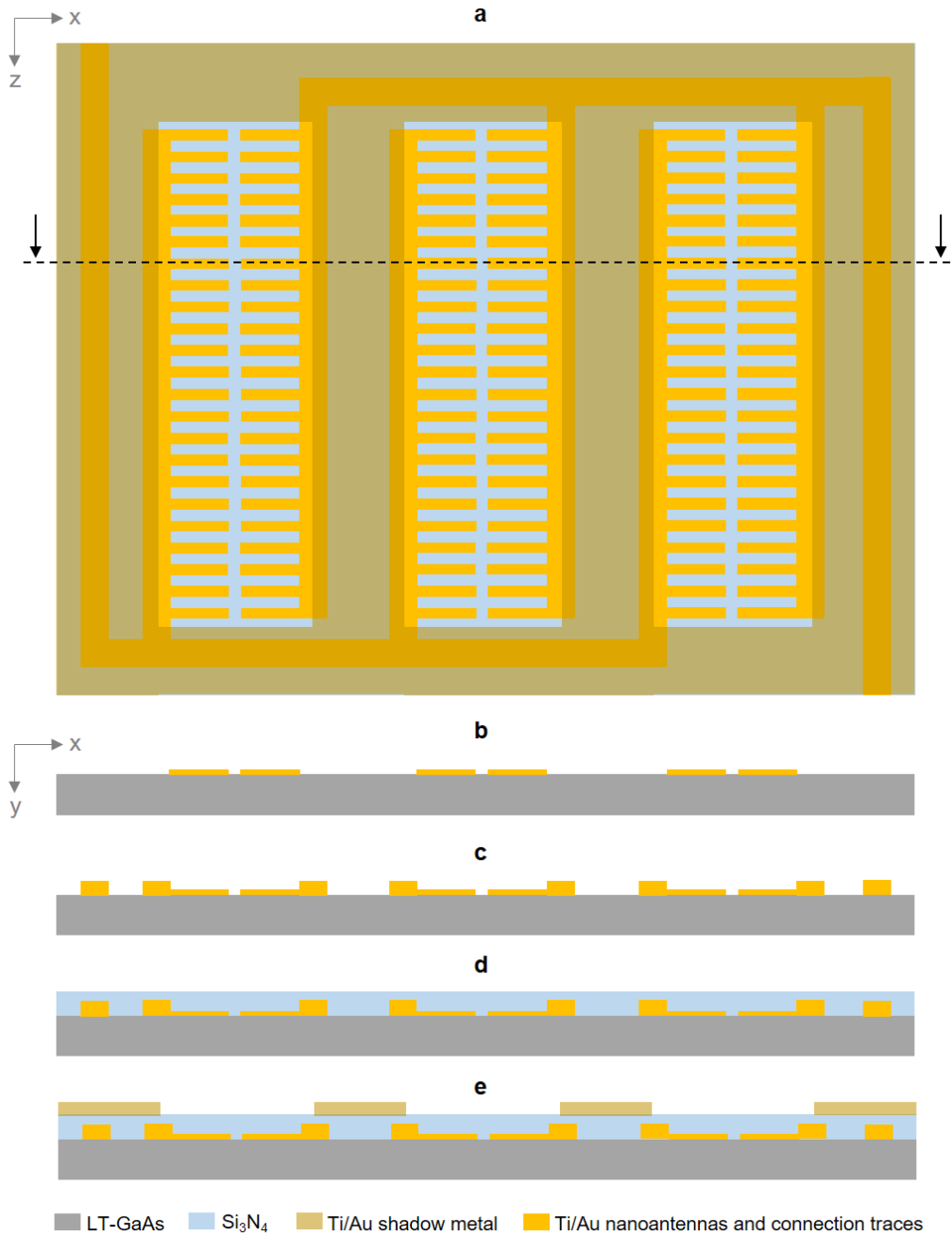

**Figure S2** Fabrication process of the THz-FPA. **a,** The top-view of one cluster composed of plasmonic nanoantenna arrays. The fabrication process steps at the cross-section marked by the dashed line are illustrated in **b**-**e**. **b,** Starting with a low-temperature-grown GaAs (LT-GaAs) substrate and patterning the plasmonic nanoantennas using electron-beam lithography, followed by metal deposition (3/47-nm Ti/Au) and lift-off. **c,** Patterning connection traces and contact pads through photolithography, followed by metal deposition (20/180-nm Ti/Au) and lift-off. **d,** Depositing a 290-nm-thick silicon nitride anti-reflection coating through plasma-enhanced chemical vapor deposition. **e,** Patterning shadow metals through photolithography, followed by metal deposition (10/90-nm Ti/Au) and lift-off.

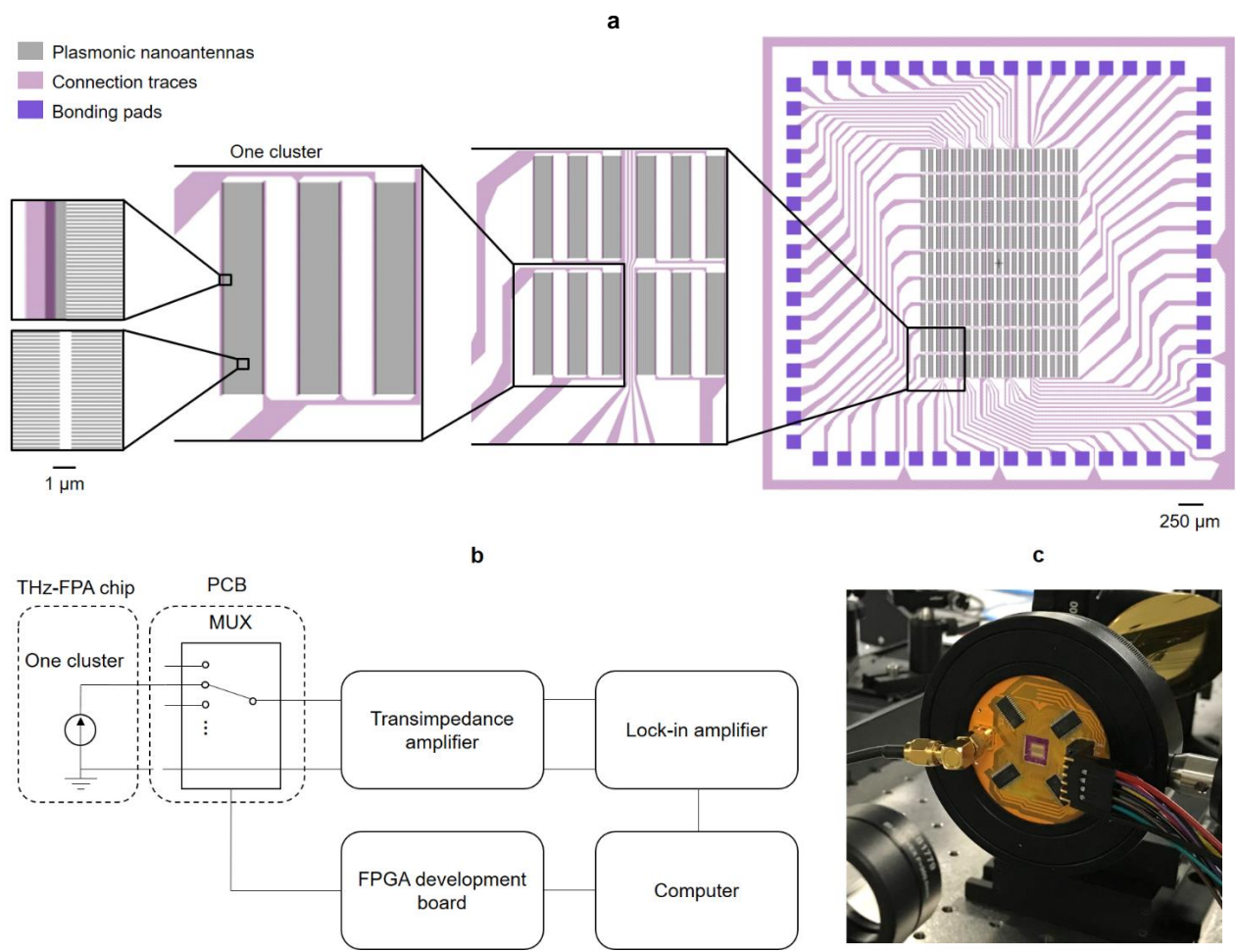

**Figure S3 a,** Layout of the THz-FPA, showing the plasmonic nanoantennas, connection traces, and bonding pads. A shadow metal layer (not shown in this layout) covers all the connection traces such that only the plasmonic nanoantennas are optically illuminated. **b**, Block diagram of the data acquisition system, which consists of an FPGA development board controlling four 16-channel multiplexers to route the FPA outputs to a transimpedance amplifier and lock-in amplifier sequentially. **c,** Photograph of the packaged THz-FPA in the imaging setup.

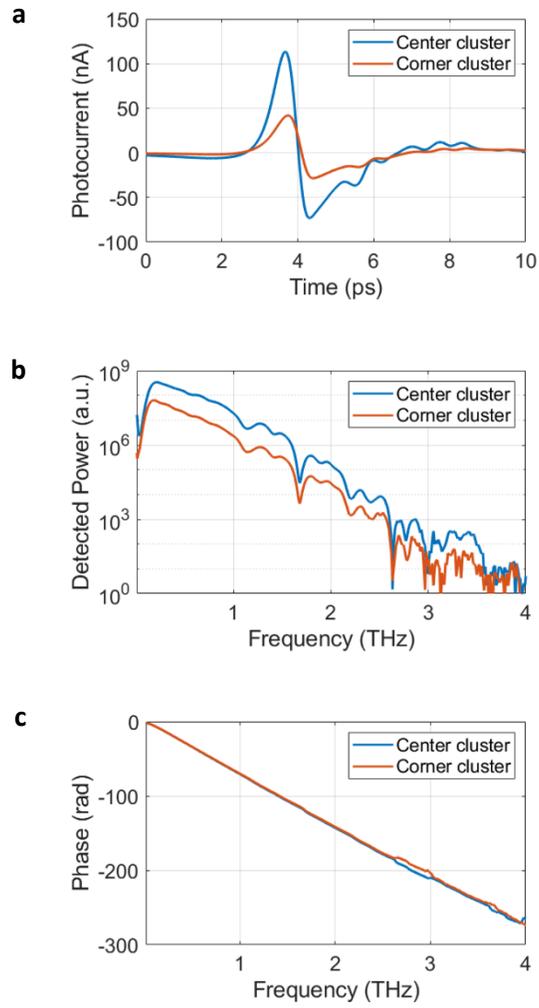

**Figure S4** The time-domain electric field, frequency-domain power spectra, and frequency-domain unwrapped phase measured at a center pixel and a corner pixel of the THz-FPA are shown in **a**, **b**, and **c** respectively.

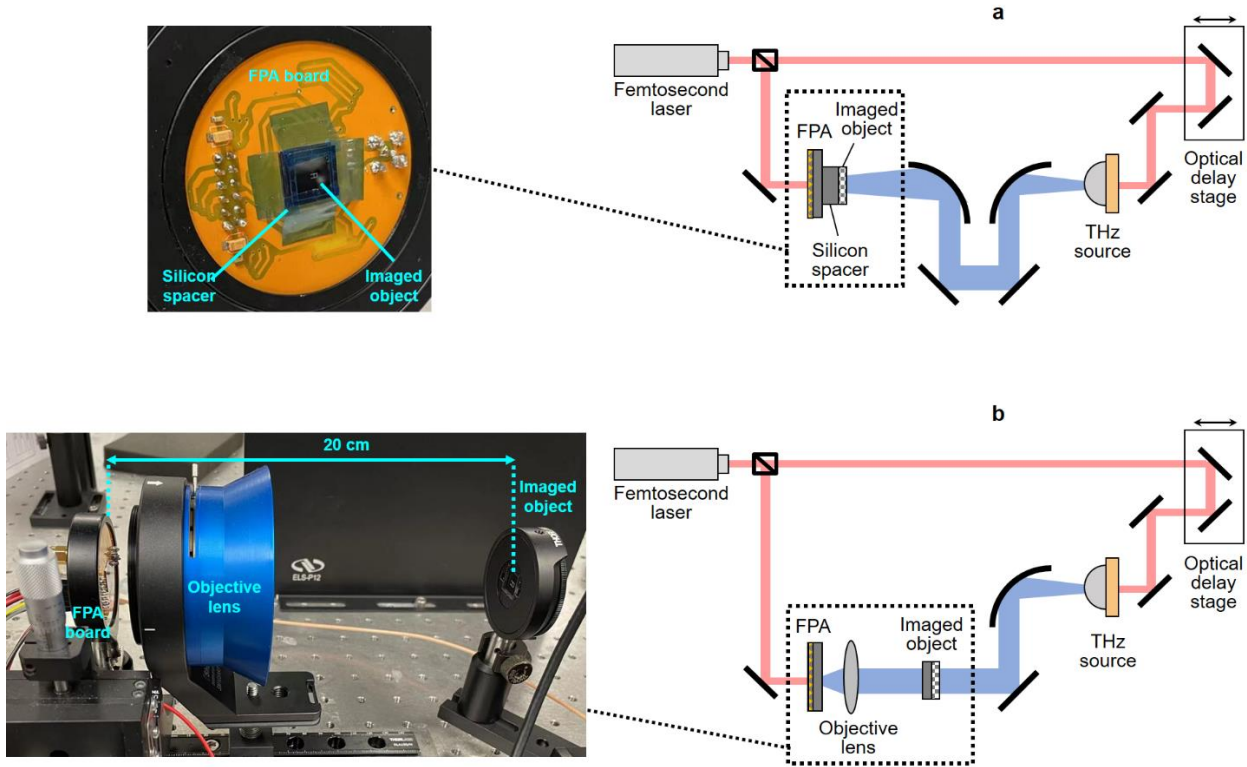

**Figure S5** Block diagrams of two experimental setups used for terahertz time-domain imaging using our THz-FPA. **a,** The imaged objects are placed at a 1.1 mm axial distance from the active area of the THz-FPA using a high resistivity silicon substrate, which serves as a spacer. The 1.1 mm distance translates to 9.4 effective wavelengths at the median wavelength of 398 μm. Two parabolic mirrors are used to collimate and focus the terahertz radiation onto the THz-FPA after interacting with the imaged object. **b,** The imaged objects are placed at a 20 cm axial distance from the active area of the THz-FPA. One parabolic mirror and one objective lens (TeraLens, Lytid) are used to collimate and focus the terahertz radiation onto the THz-FPA after interacting with the imaged objects while providing a demagnification factor of ~2.76.

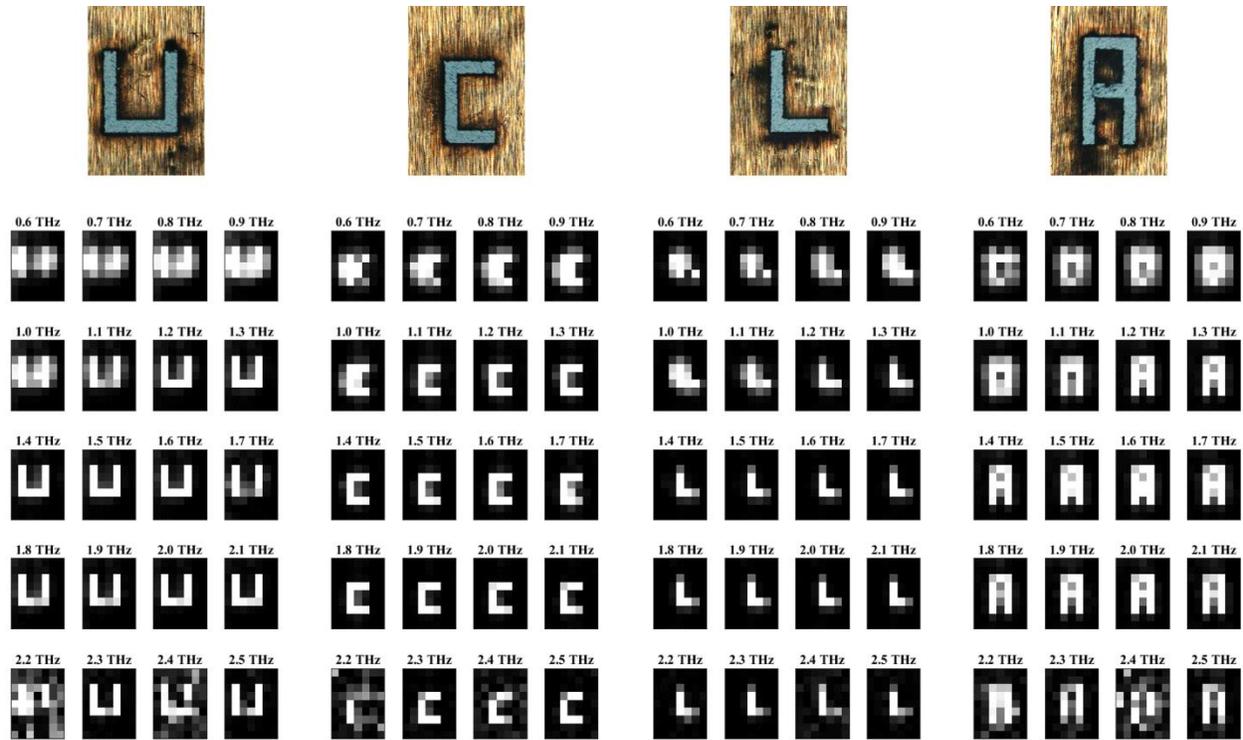

**Figure S6** Raw spectral images of 'U', 'C', 'L', and 'A' patterns created in a 65 μm-thick copper plate captured by the fabricated THz-FPA over a 0.6-2.5 THz frequency range [88]. The minimum aperture width of all the patterns is 160 μm. The image resolution is degraded at lower frequencies due to the diffraction limit. The image quality also degrades at frequencies close to the water vapor absorption lines, 1.1 THz, 1.7 THz, 2.2 THz, and 2.4 THz, due to the high signal attenuation.

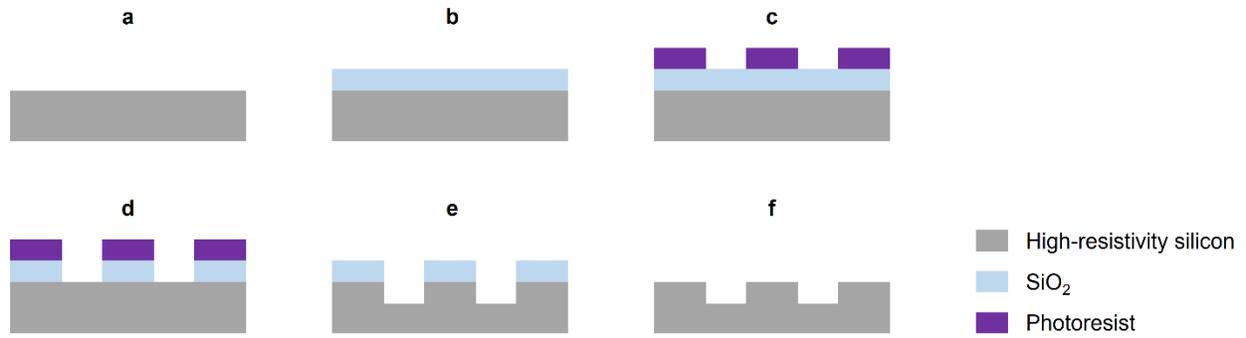

**Figure S7** Fabrication process of the phase imaging objects. **a,** Starting with a high-resistivity silicon substrate. **b,** Growing a thermal oxide layer on the substrate. **c,** Defining the image patterns through photolithography. **d,** Forming a hard etch mask by dry etching the silicon dioxide layer. **e,** Forming the final image patterns by etching silicon through deep reactive-ion etching. **f,** Removing the silicon dioxide hard mask through buffered oxide etching.

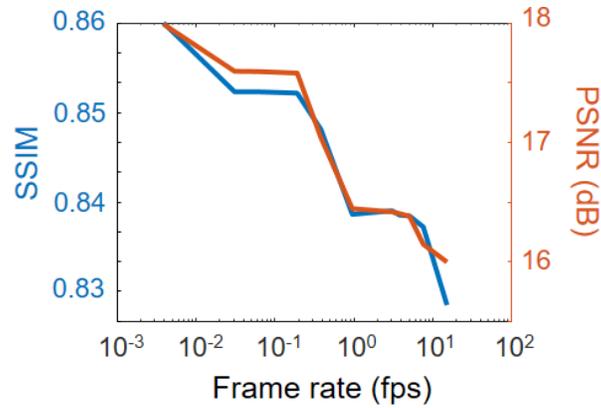

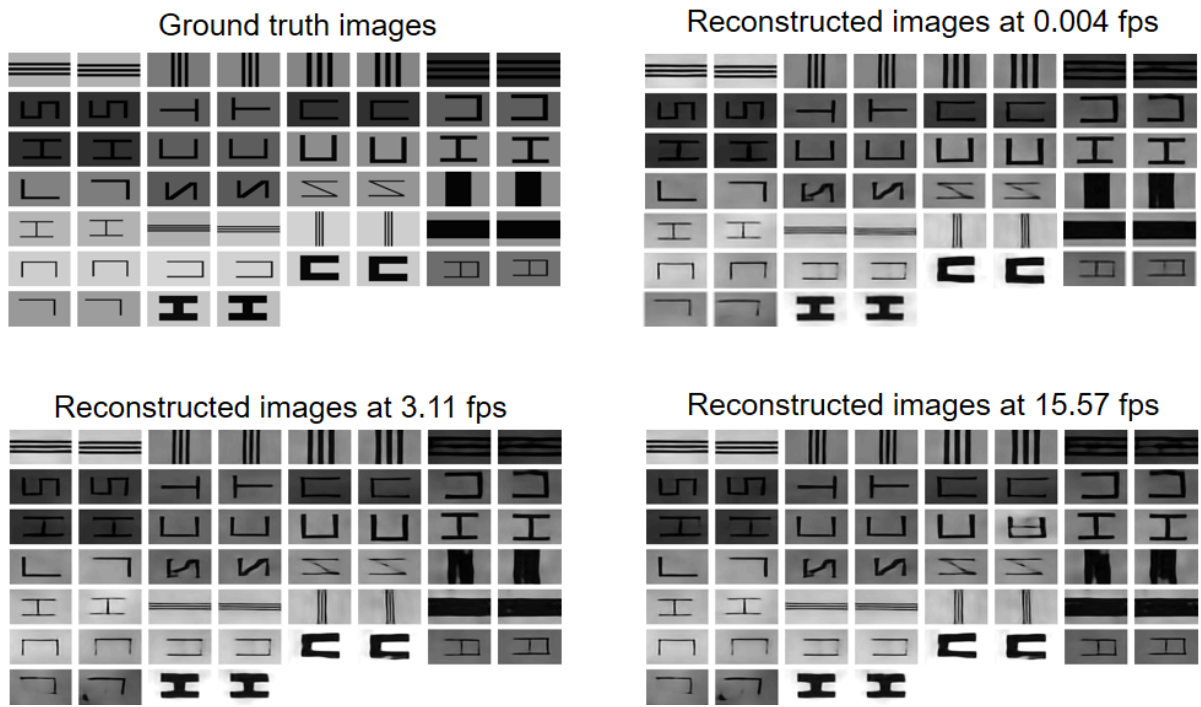

**Figure S8** The speed of the THz-FPA imaging system for capturing the time-domain data from all the plasmonic clusters depends on (1) the speed and range of the optical delay stage used to vary the optical path difference between the optical pump and terahertz beams incident on the THz-FPA, (2) the number of the time-domain traces that are captured and averaged to resolve each image, and (3) the data acquisition scheme. By setting the speed and range of the optical delay stage to 50 mm/s and 25 mm, respectively, and capturing 8 time-domain traces for each image, we achieve an imaging speed of 0.004 fps. In this setting, one time-domain trace from each selected cluster is fully captured while moving the optical delay stage over a 25 mm range, before selecting another cluster and repeating the same data capture. The imaging speed can be significantly increased if the time-domain data from all clusters are captured sequentially while moving the optical delay stage over a 25 mm range. Through this data acquisition approach, the frame-rate is increased to 3.11 fps when reducing the optical delay stage range to 2.5 mm and capturing 5 time-domain traces for each image. Under this condition, reducing the number of the captured time-domain traces to 1 further increases the frame-rate to 15.57 fps. The depth and shape of all imaged objects are successfully resolved when effectively increasing the frame-rate from 0.004 fps to 15.57 fps with a small reduction in the image reconstruction SSIM (from 0.86 to 0.83) and PSNR (from 18 to 16 dB). The capability of the THz-FPA in offering fast imaging

speeds enabled us to capture a dynamic video of water flow in adjacent plastic pipes at 16 fps (see the Supplementary Video).

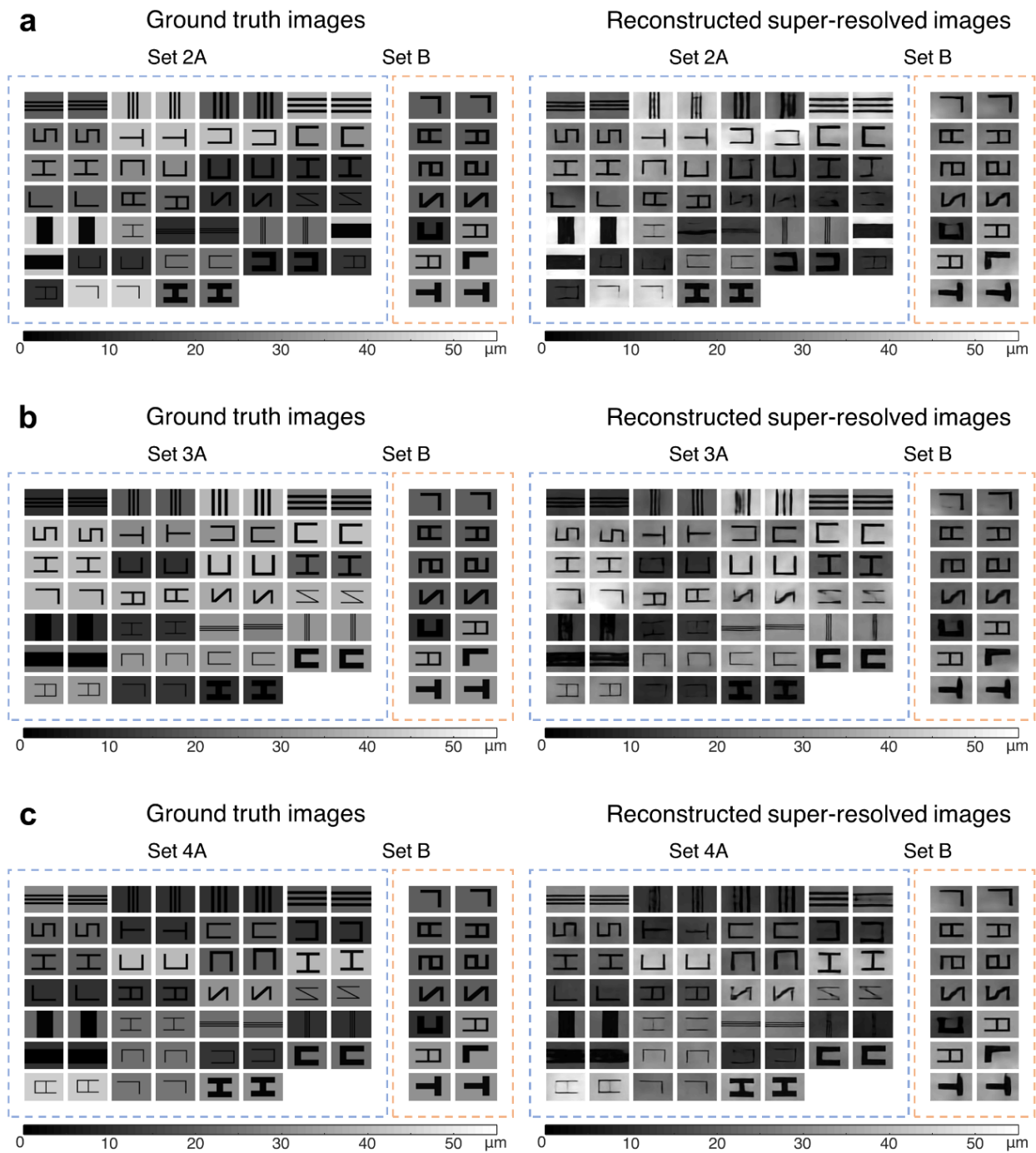

**Figure S9** Generalization performance of the presented PSR-enhanced THz-FPA on different test sets: Sets 2A, 3A, and 4A shown in **a**, **b**, and **c**, respectively. The CNN architecture shown in Fig. 2 of the main text was trained from scratch to test on these 3 different datasets, where the training/testing partition of the image data is different for each case (see Methods section of the main text).

Ground truth images

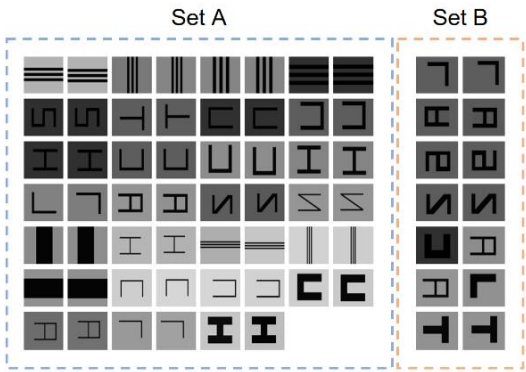
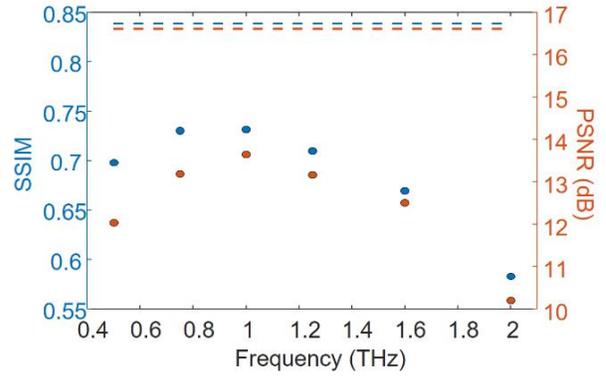
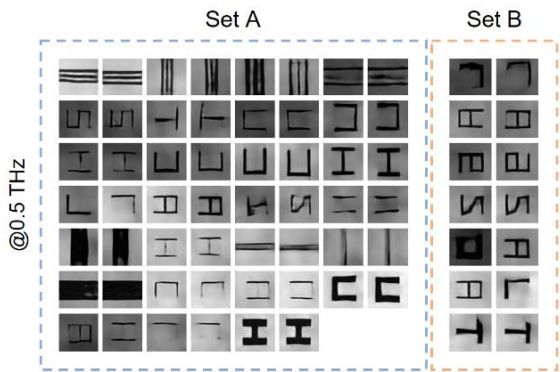
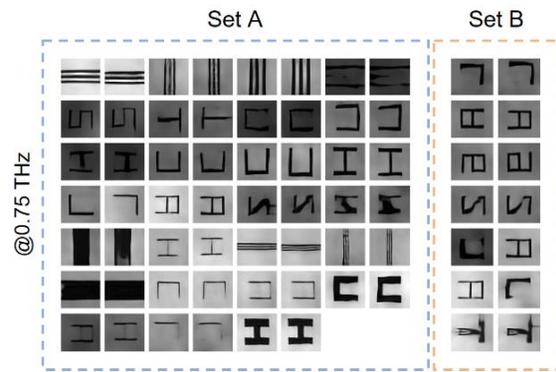
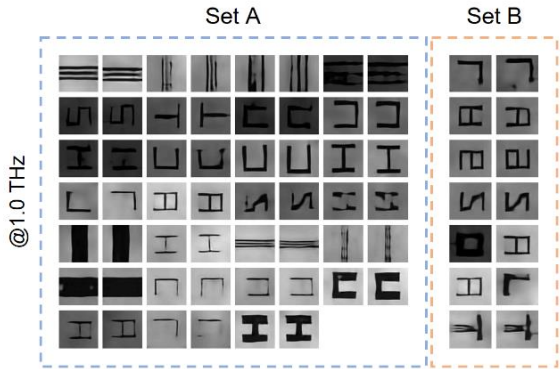
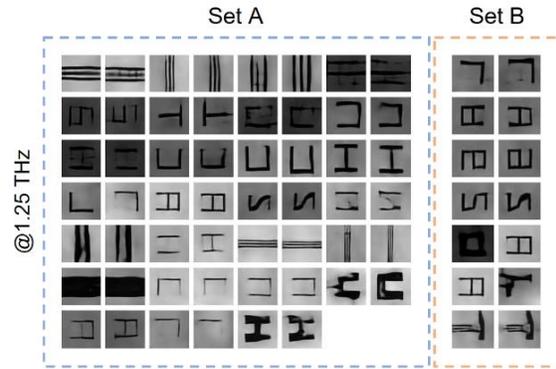
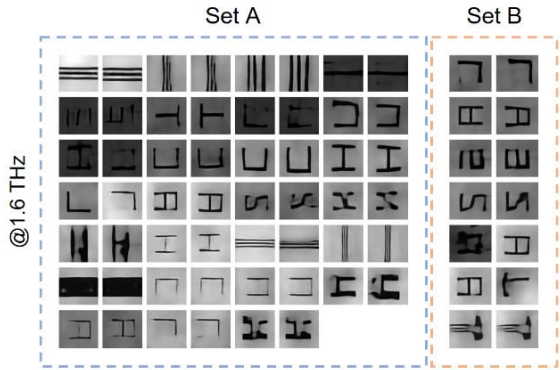
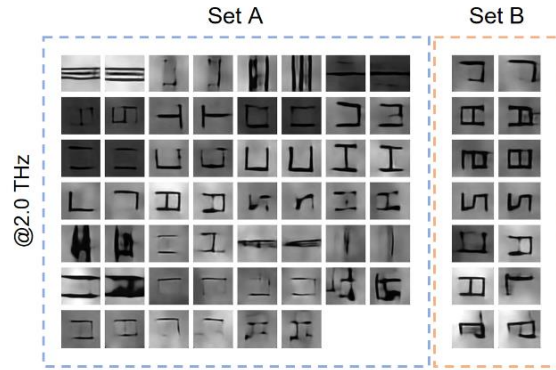

**Figure S10** Single-frequency super-resolution networks and their comparison with our multi-spectral PSR-enhanced THz-FPA results. Ground-truth thickness images in the test set that are identical to the ones shown in Fig. 2 of the main text are shown on the top-left. There are six CNNs separately trained to perform PSR solely based on the amplitude and phase channel of a single spectral component at, 0.5 THz, 0.75 THz, 1.0 THz, 1.25 THz, 1.6 THz and 1.8 THz. The SSIM (blue) and PSNR (red) values achieved by these single-frequency PSR networks are depicted in the graph shown on the top-right, whereas the dashed lines represent the corresponding values attained by the multi-spectral PSR-enhanced THz-FPA results presented in the Fig. 2 of the main text.

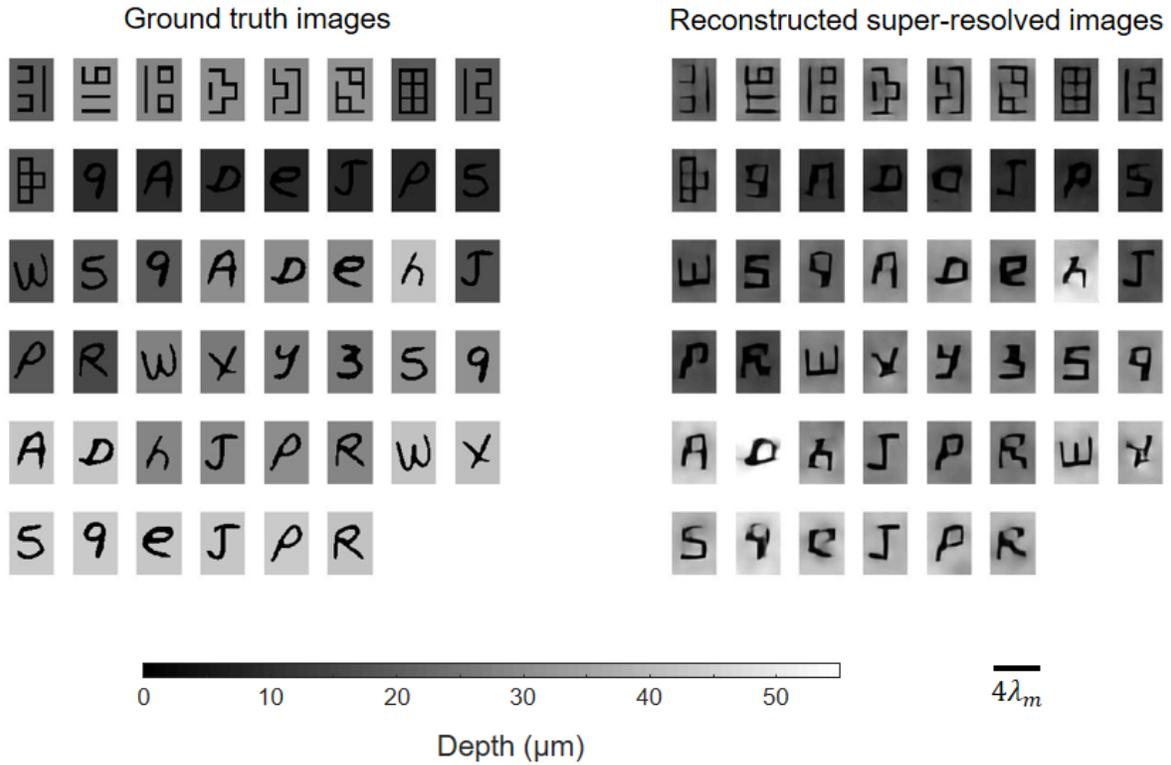

**Figure S11** The ground truth and the reconstructed images of complex and irregular patterns, including hand-written numbers and letters, that were fabricated and etched in silicon. These objects are blindly tested with the same PSR neural network used in Fig. 2. No patterns with similar shapes were seen by the neural network during its training. The depth and shape of the patterns are successfully reconstructed with an average PSNR of 12 dB and SSIM of 0.64, confirming the generalization performance of the presented PSR-enhanced THz-FPA.

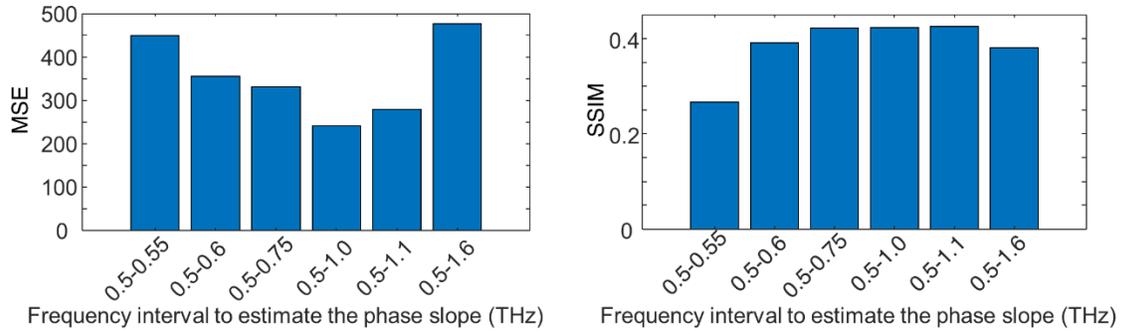

**Figure S12** The impact of the spectral bandwidth on the quality of the thickness image estimate. Mean-square-error (MSE) (left) and Structural Similarity Index Measure (SSIM) (right) of the thickness image recovered using the slope of the unwrapped phase of the spectral components within different bands. Note that these metrics are not calculated based on the output of the PSR network, unlike e.g., the SSIM metrics reported in Fig. 4 of the main text. The band from 0.5 THz to 1 THz was found to be optimal, since it provides the lowest MSE. The reason behind the increasing MSE with the inclusion of the components beyond 1 THz is the existence of water absorption lines.

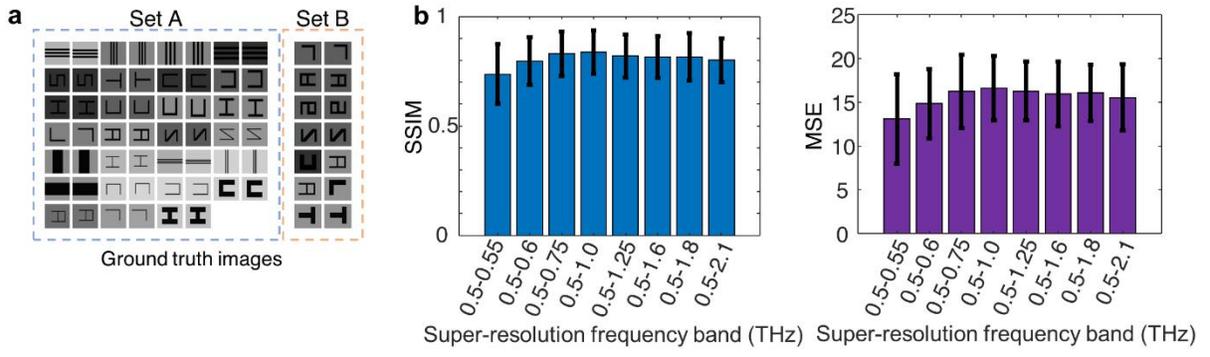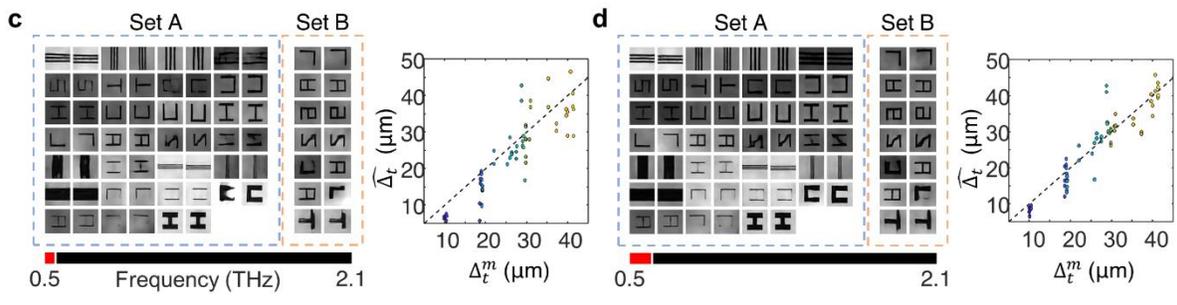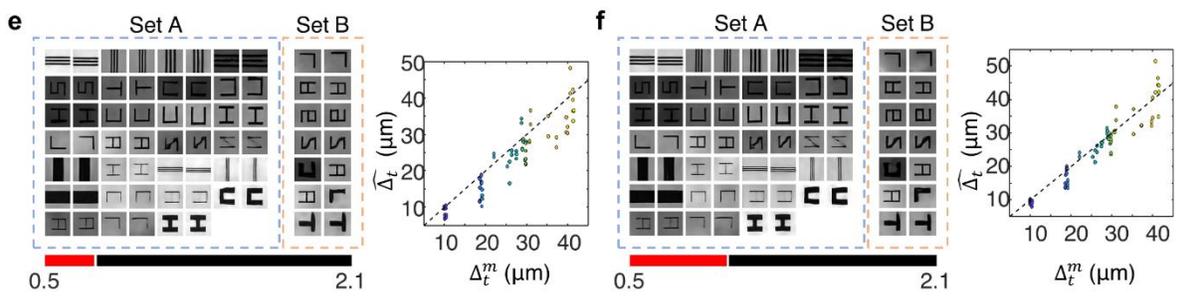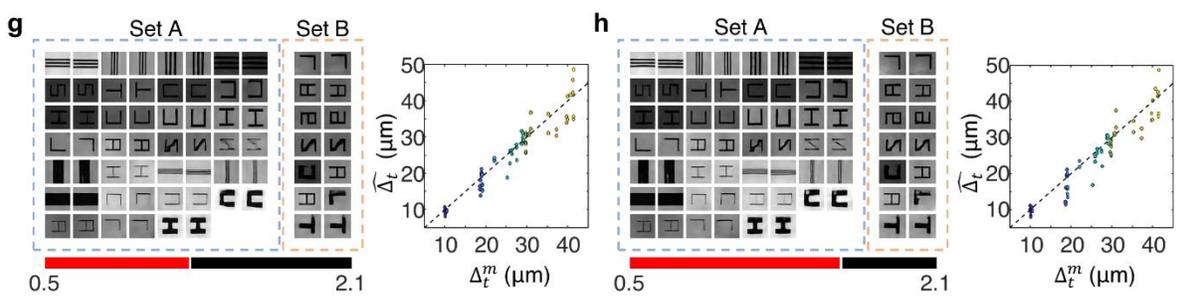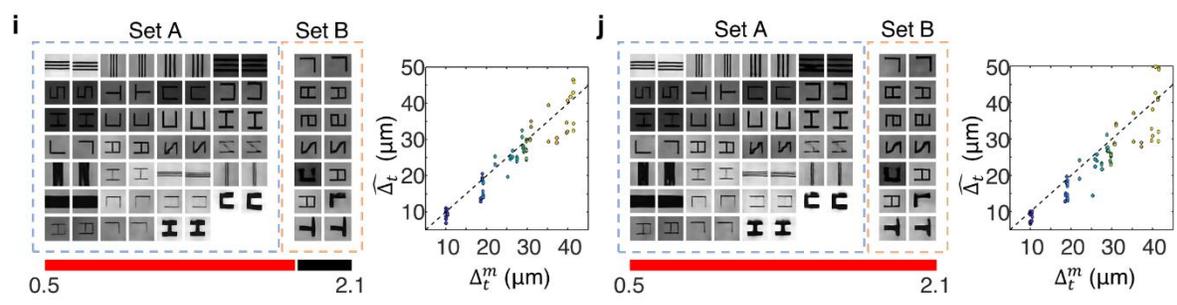

**Figure S13** The impact of the spectral bandwidth on the presented PSR network inference. **a,** Ground-truth thickness images in the test set that is identical to the one shown in Fig. 2 of the main text. Eight different CNNs were separately trained and tested on this test set each performing PSR based on a different spectral band. **b,** The SSIM and PSNR of the super-resolved images illustrated in c-j. **c-j,** The super-resolved images at the output layer of the PSR networks (left) and the associated thickness contrast estimation (right) when the PSR network was trained and tested based on the spectral components within the bands (c) 0.5 THz – 0.55 THz, (d) 0.5 THz – 0.6 THz, (e) 0.5 THz – 0.75 THz, (f) 0.5 THz – 1.0 THz, (g) 0.5 THz – 1.25 THz, (h) 0.5 THz – 1.6 THz, (i) 0.5 THz – 1.8 THz and (j) 0.5 THz – 2.1 THz. The super-resolved images shown in (f) are identical to the ones shown in Fig. 2 of the main text.

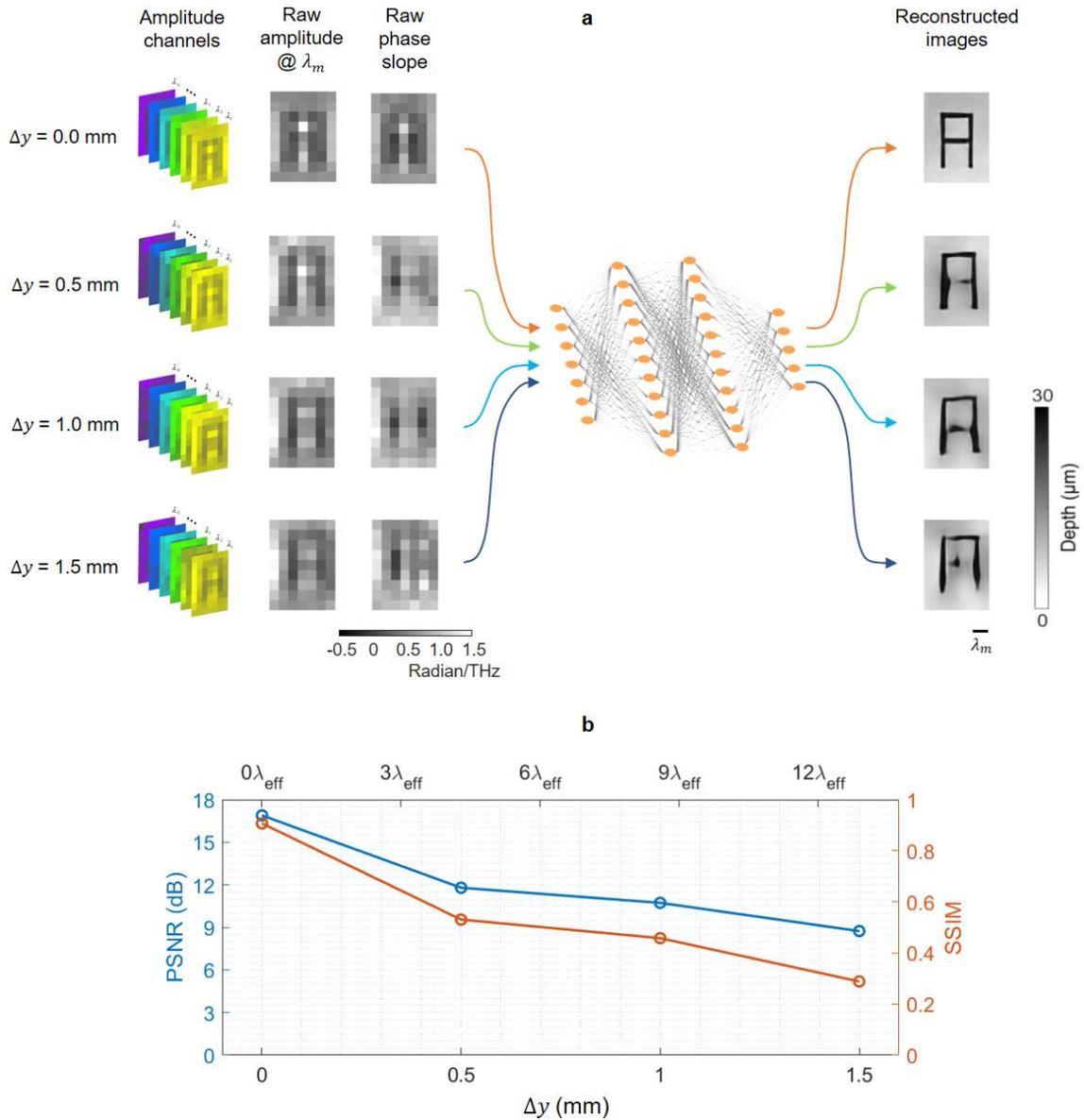

**Figure S14** Depth-of-field analysis for the PSR-enhanced THz-FPA imaging system (shown in Fig. S5a). By increasing the number of silicon spacers placed between the imaged object and the THz-FPA, the axial distance of the object from the THz-FPA is experimentally increased by different Δy values. **a**, Reconstructed images for different axial displacements Δy = 0, 0.5, 1, and 1.5 mm. Although all the neural network training is performed for imaged objects placed at the 1.1 mm axial distance, the shape and depth of the object are successfully reconstructed for axial displacements as large as 1.5 mm (which is equivalent to 12.8 effective wavelengths at the median frequency of 0.754 THz). **b**, The PSNR and SSIM of the reconstructed images at different axial displacements.

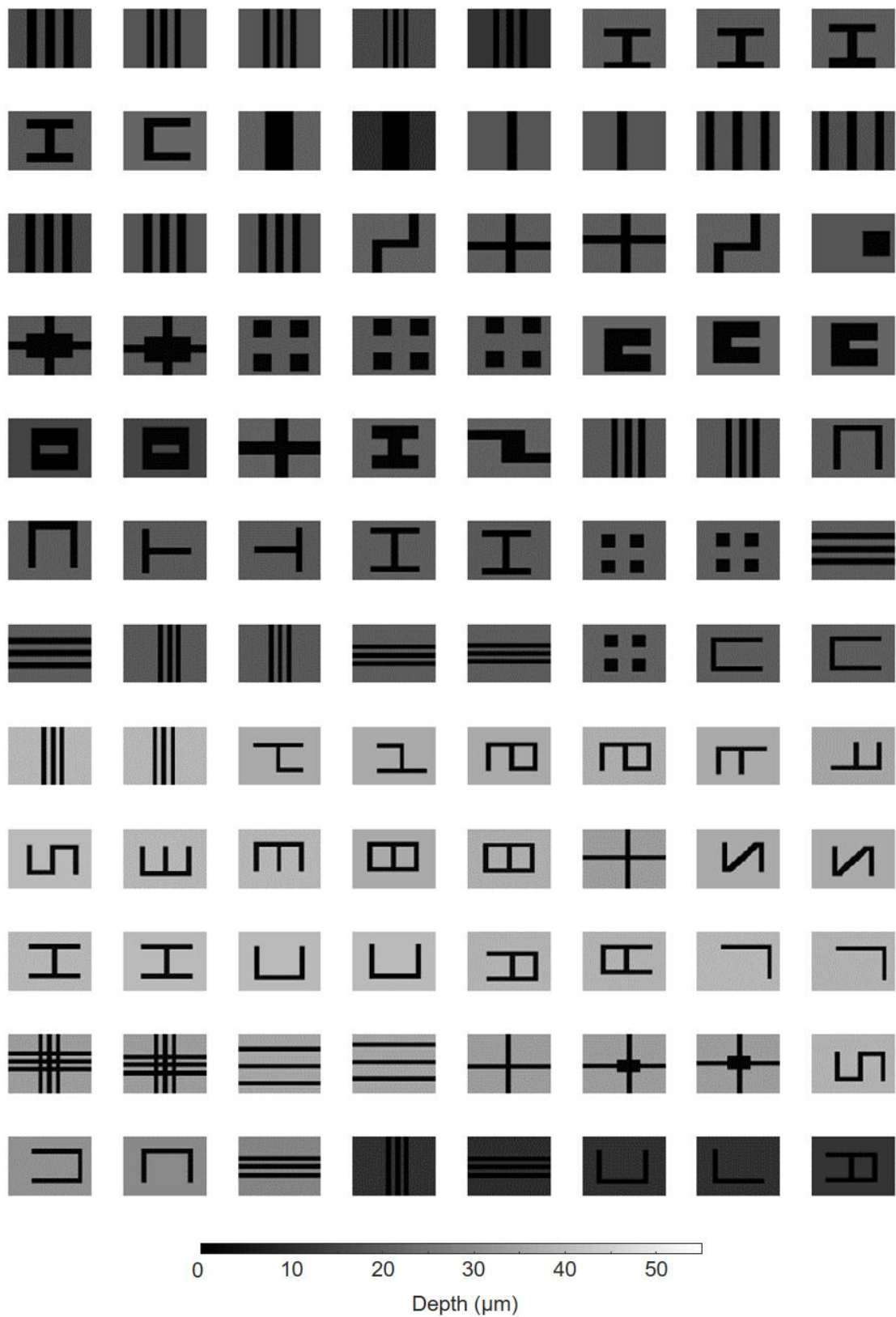

**Figure S15 (1/3)** Ground truth thickness images of the training set for the PSR neural network reported in Fig. 2.

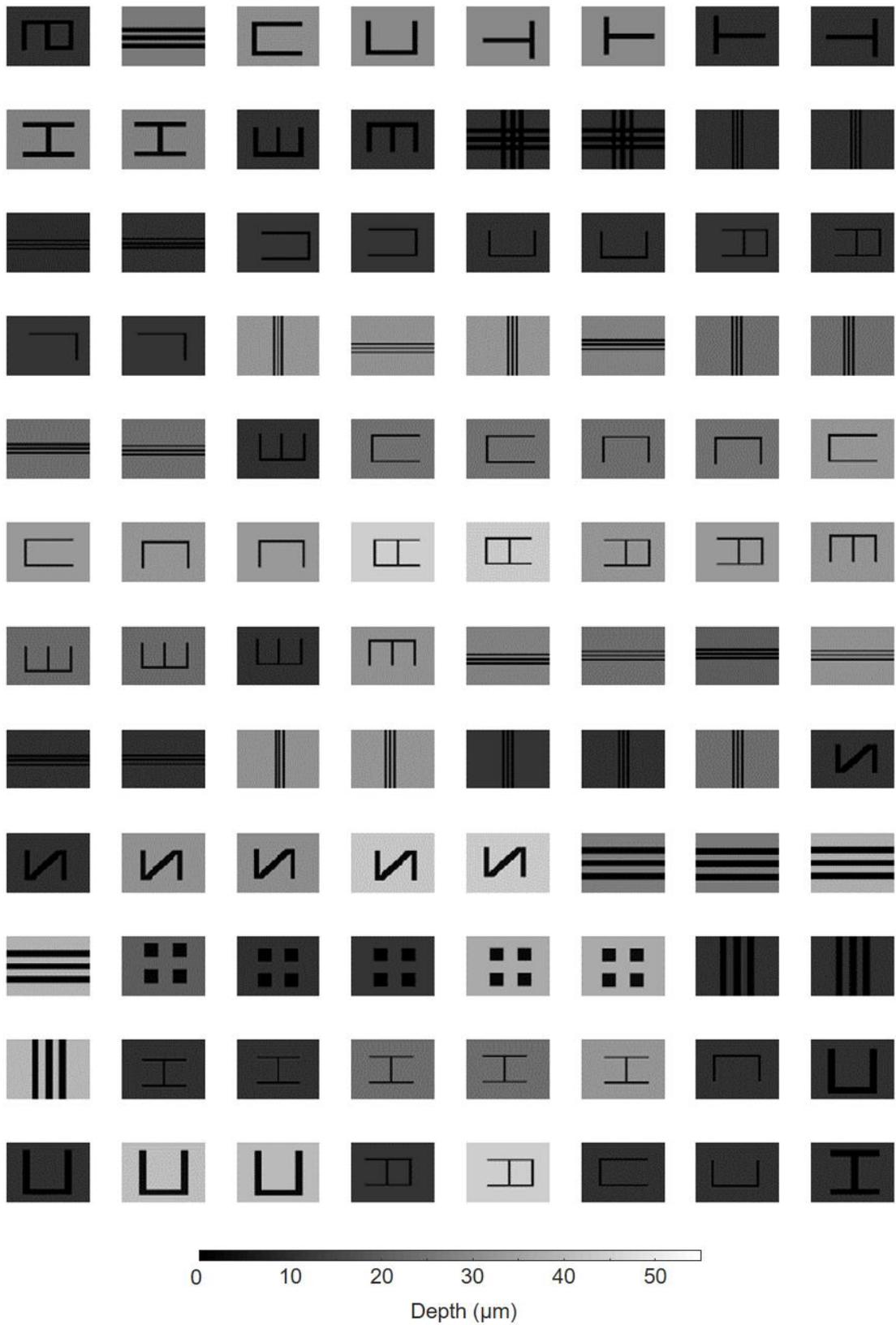

**Figure S15 (2/3)** Ground truth thickness images of the training set for the PSR neural network reported in Fig. 2.

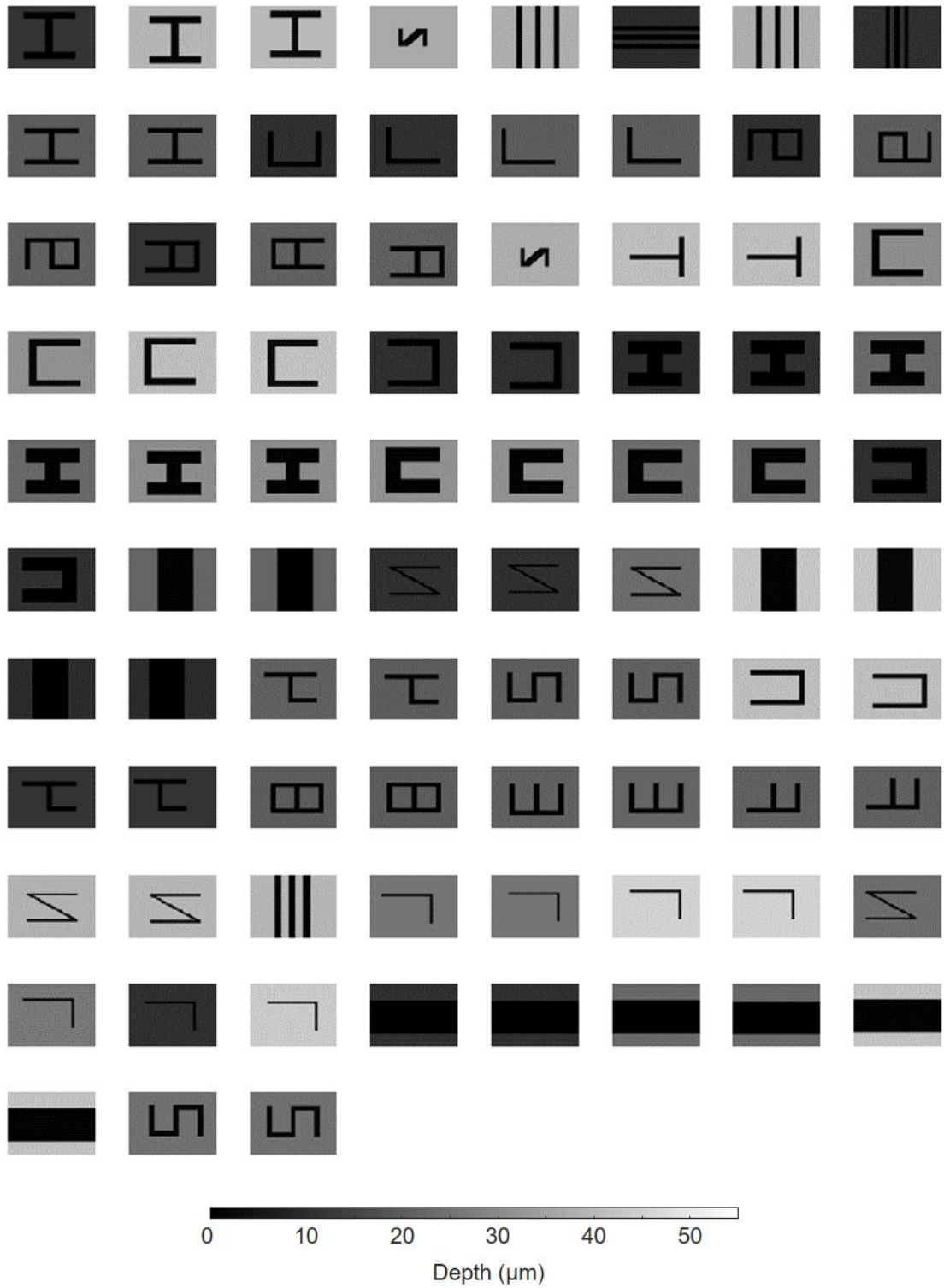

**Figure S15 (3/3)** Ground truth thickness images of the training set for the PSR neural network reported in Fig. 2.

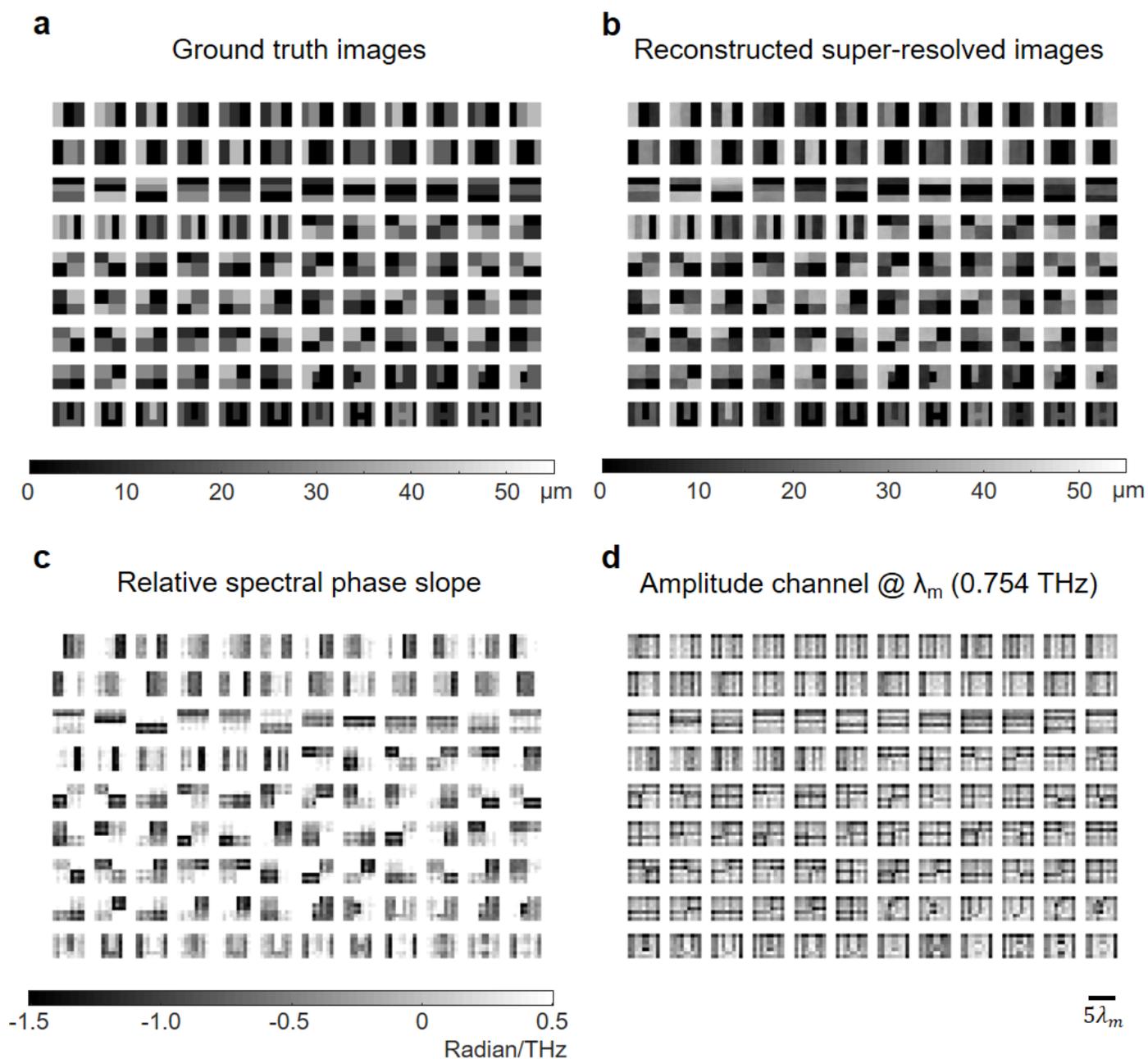

**Figure S16** PSR-enhanced THz-FPA imaging with the capability of resolving objects with non-binary depths and smaller contrast. The original PSR neural network is enhanced by transfer learning using synthetic image data generated by a terahertz forward model (see the Supplementary Methods). **a,** Ground truth thickness images of the tested objects. **b,** Super-resolved images using the transfer-learned-PSR-enhanced THz-FPA. **c,** The relative spectral phase slope of the objects computed based on the slope of the unwrapped spectral phase distribution. **d,** The amplitude channel at the median frequency, $f_m$ = 0.754 THz. The depth and shape of the patterns are successfully reconstructed with a PSNR of 29.4 dB and SSIM of 0.93, confirming the success of the transfer-learned, PSR-enhanced THz-FPA imaging system.

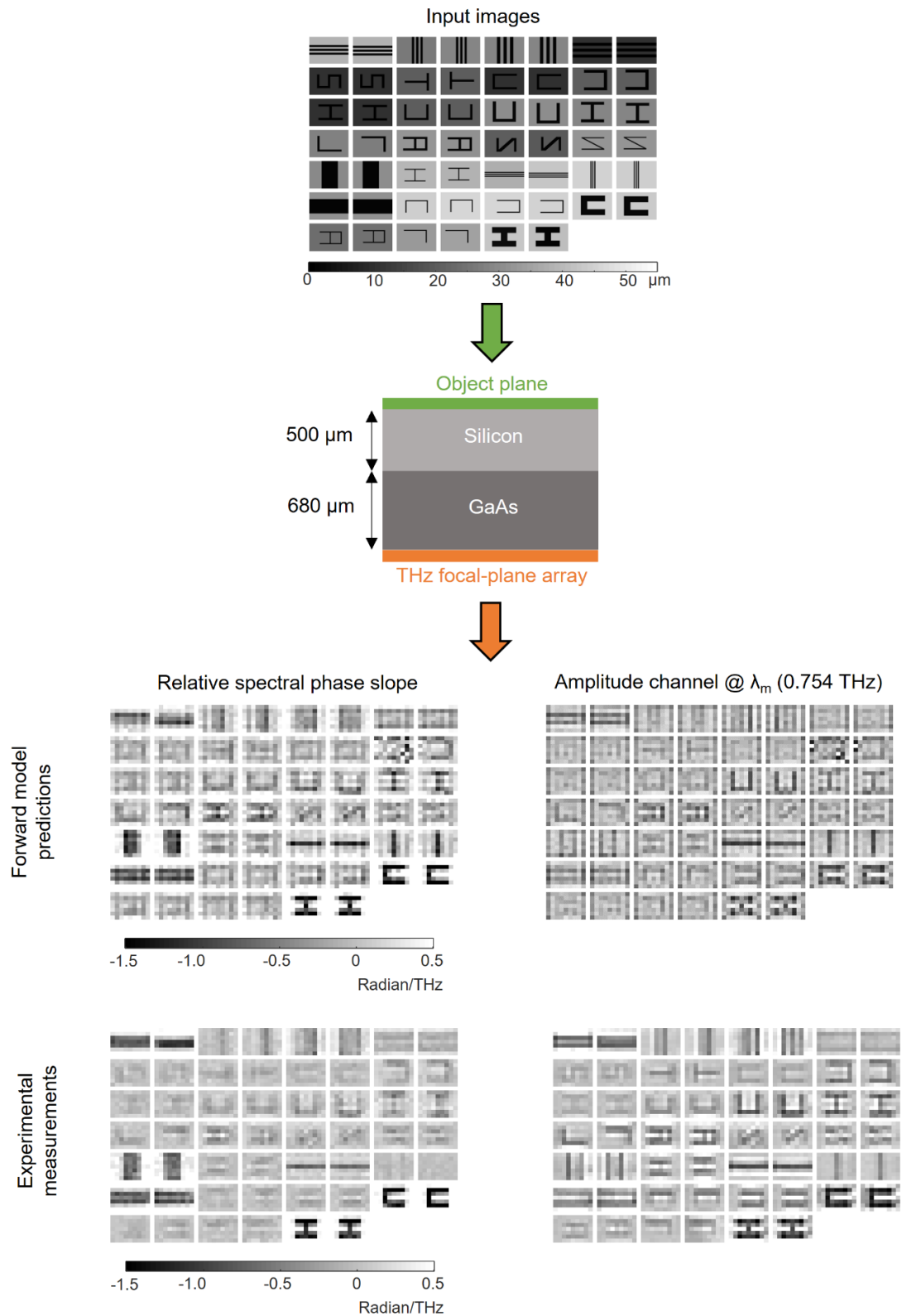

**Figure S17** Comparison of the predictions of the terahertz wave forward model against the experimental measurements, demonstrating the agreement between the generated output terahertz signals at the FPA plane and the corresponding experimental results.

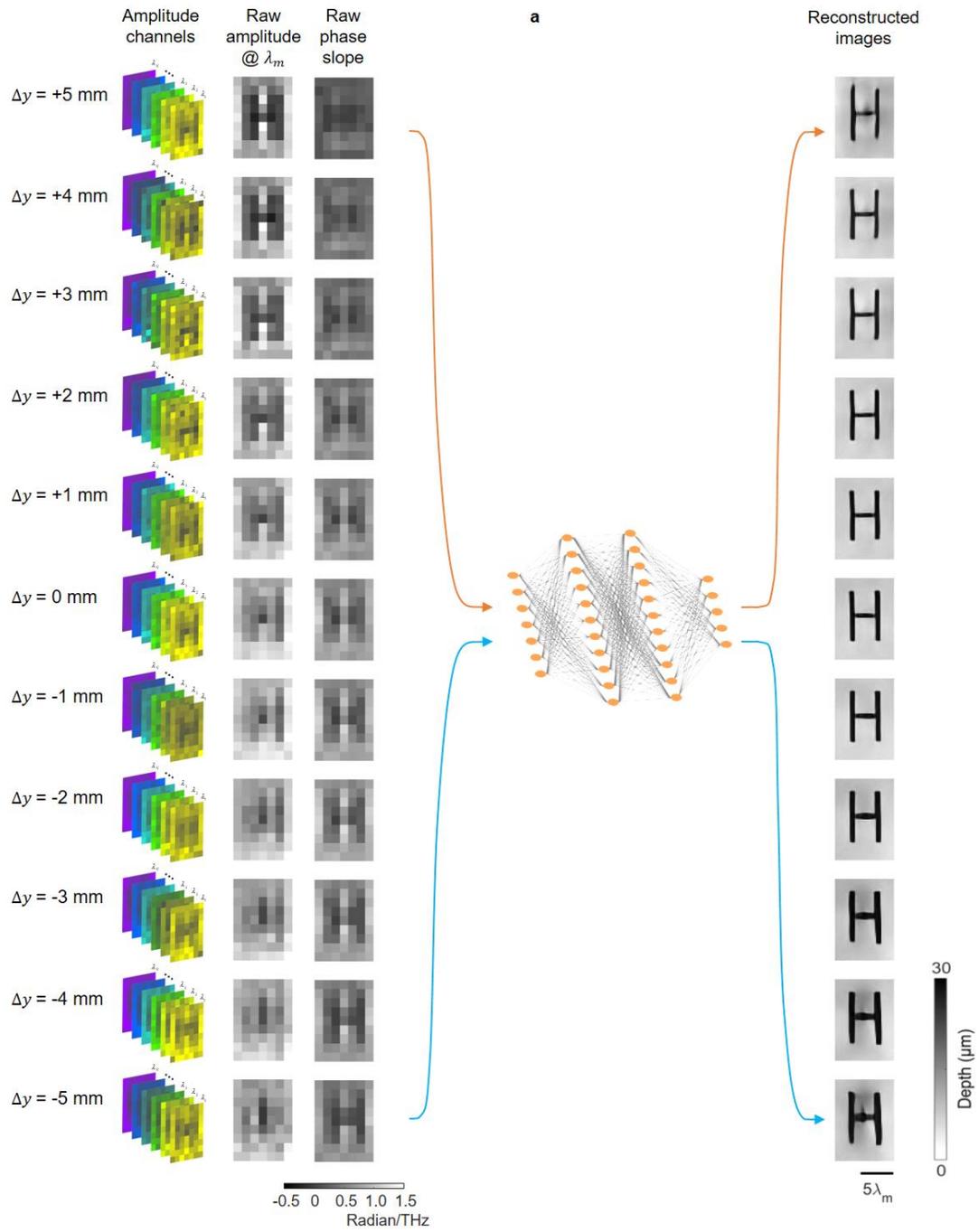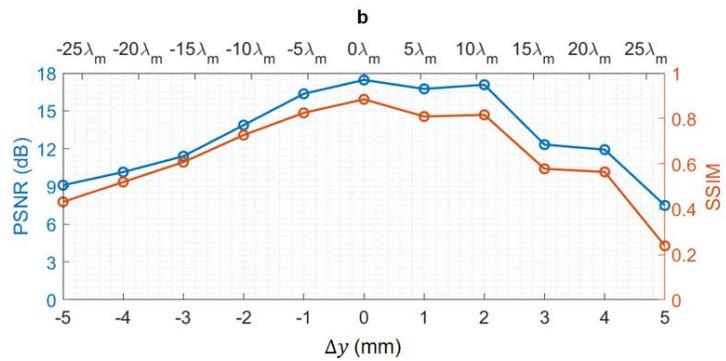

**Figure S18** Depth-of-field analysis for the PSR-enhanced THz-FPA imaging system (shown in Fig. S5b). **a**, Reconstructed images for different axial experimental displacements, $\Delta y$, ranging from -5 mm to 5 mm. Although all the neural network training is performed for imaged objects placed at the 20 cm axial distance, the shape and depth of the object are successfully reconstructed for axial displacements as large as 10 mm (which corresponds to an axial distance of 53.3 wavelengths at the median frequency of 1.6 THz). **b**, The PSNR and SSIM of the reconstructed images at different axial displacements.

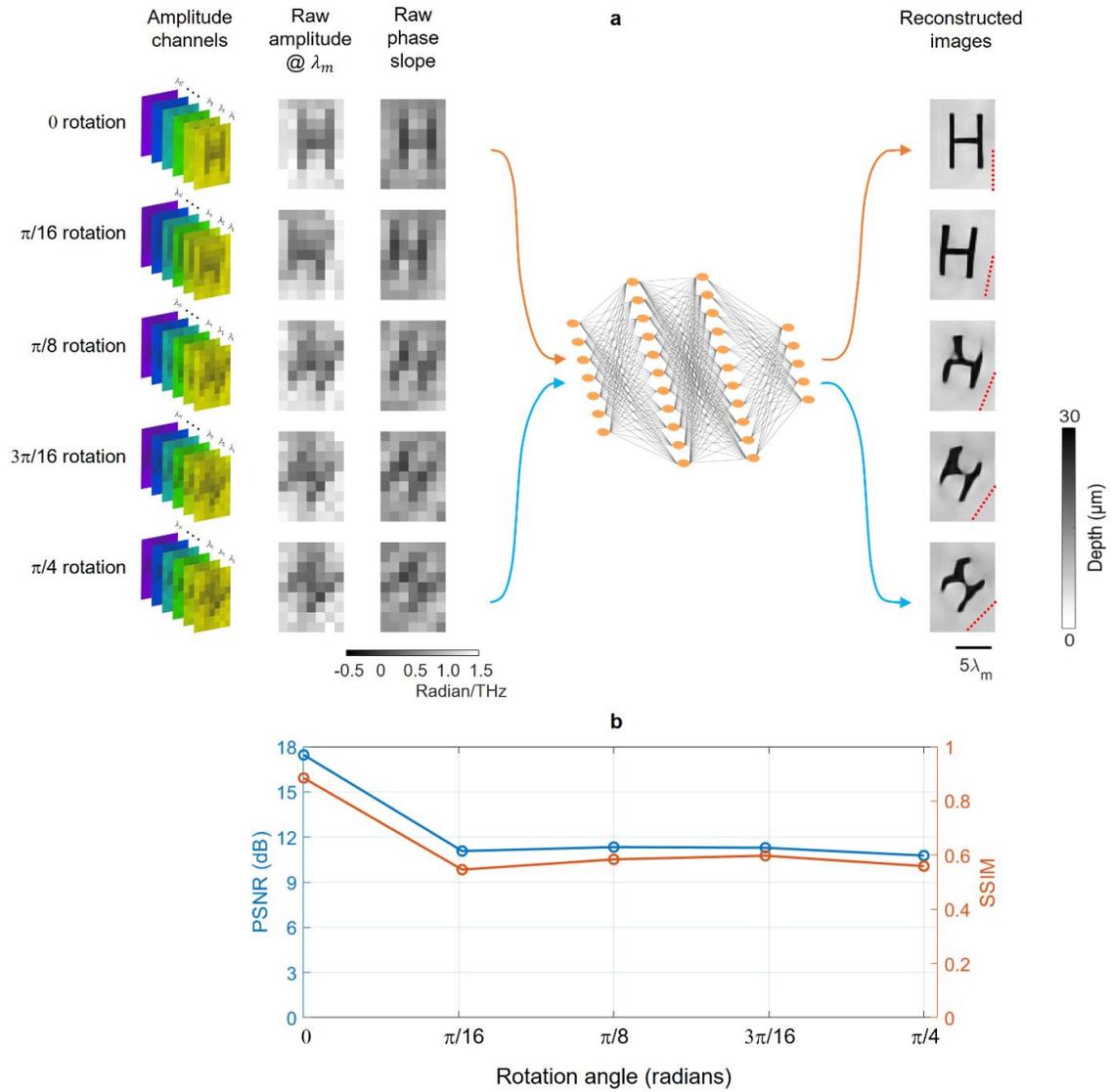

**Figure S19** Rotation sensitivity analysis for the PSR-enhanced THz-FPA imaging system (shown in Fig. S5b). **a**, Reconstructed images for different experimental rotation angles inside the object plane. Although all the neural network training is performed for imaged objects placed at a zero rotation angle, the shape and depth of the objects are successfully reconstructed for rotation angles as large as π/4. **b**, The PSNR and SSIM of the reconstructed images at different rotation angles. The dashed red lines at the bottom right side of the reconstructed images are oriented along the rotation angle of the imaged object, and clearly show the success of the imaging system in preserving the orientation information of the imaged object in these blinded neural network reconstructions.

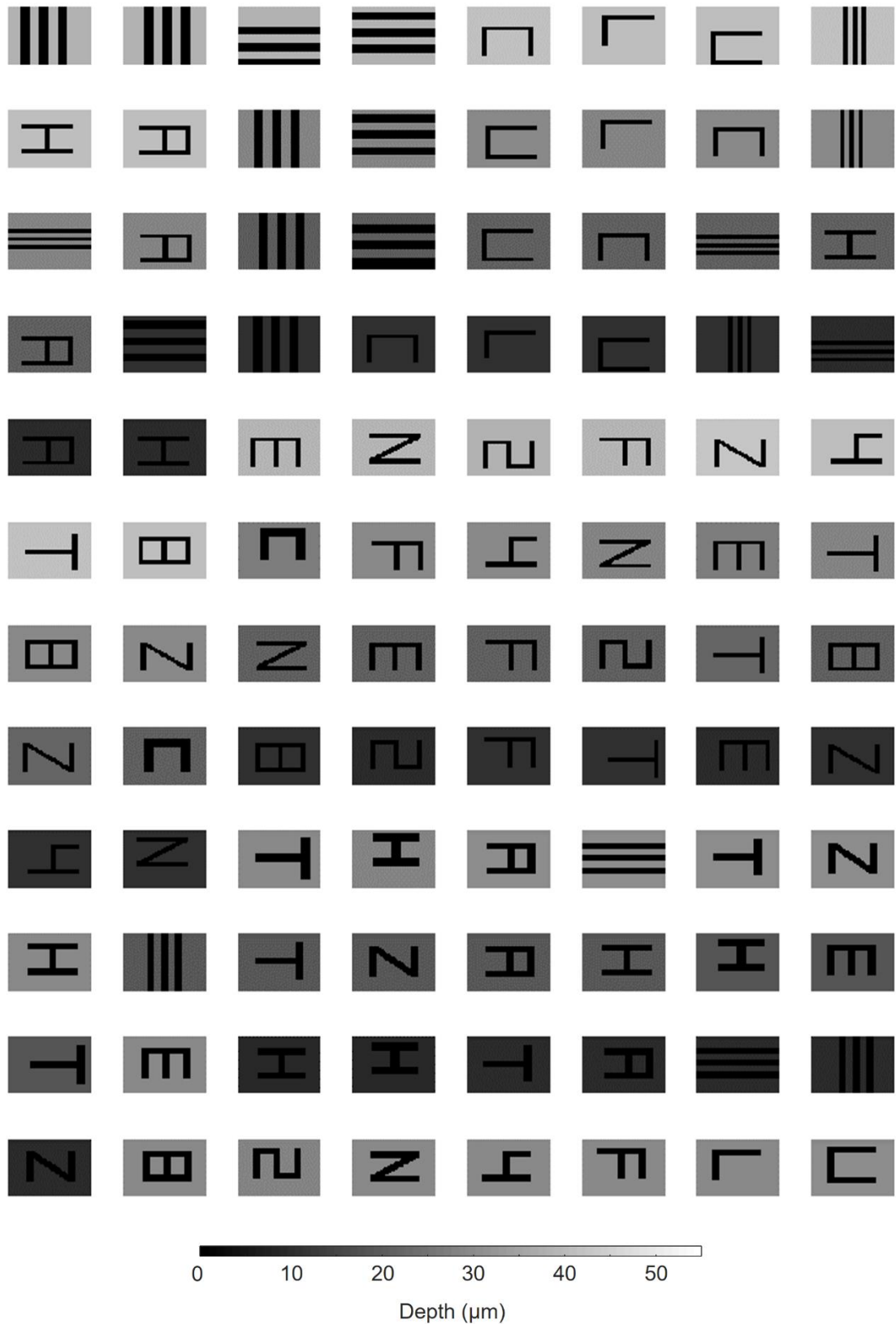

**Figure S20 (1/2)** Ground truth thickness images of the training set for the PSR neural network reported in Fig. 5.

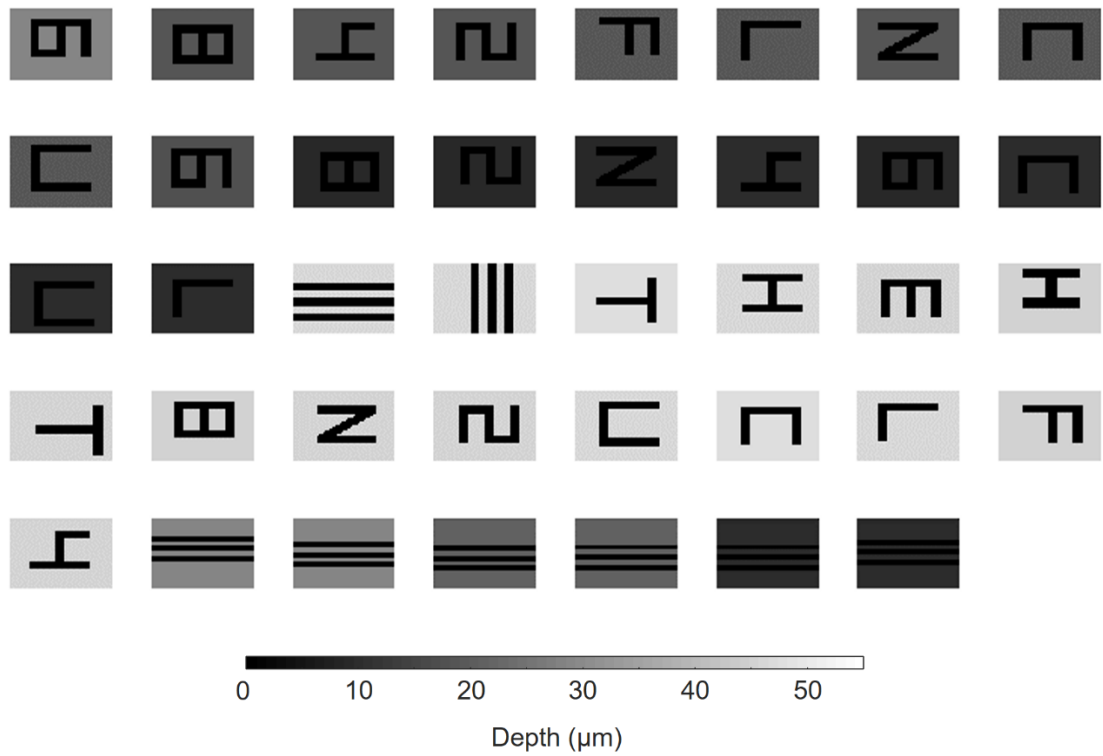

**Figure S20 (2/2)** Ground truth thickness images of the training set for the PSR neural network reported in Fig. 5.

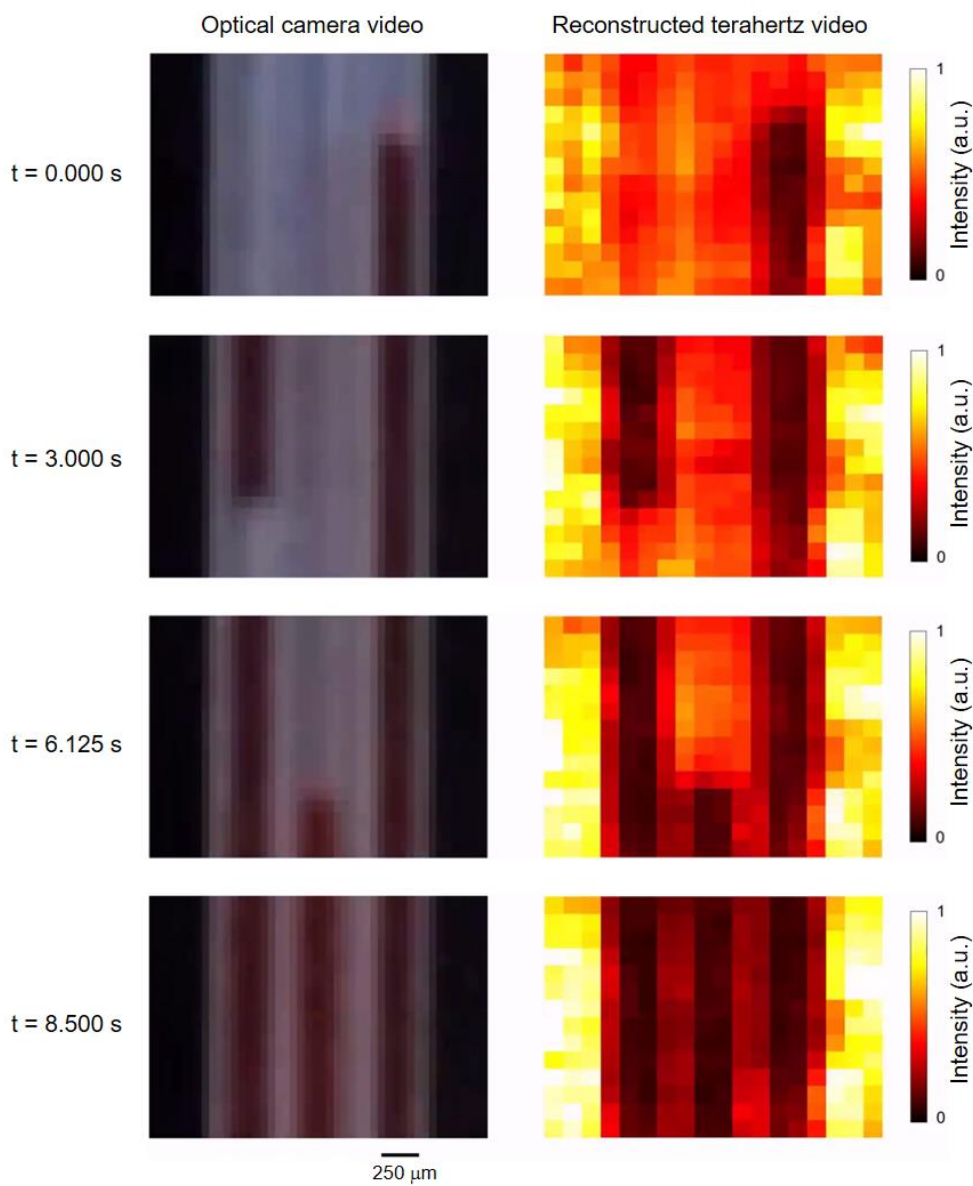

**Figure S21** Holographic reconstructions of the video frames captured by the THz-FPA from water flow in adjacent plastic pipes (see Supplementary Methods and Supplementary Video for the full video). Three plastic pipes (PE tubing, World Precision Instruments) are used in this experiment with an outer/inner diameter of 500/250 μm are glued side-by-side to the backside of the THz-FPA substrate (680-μm-thick GaAs) using a 75-μm-thick sticky tape. A syringe pump is used to move water through the pipes and terahertz video frames are captured by the THz-FPA at 16 fps (for system specifications see Supplementary Fig. S8). A color (red) dye (tie dye kit, Patifeed) is added to water to better visualize its movement and an optical camera is also used to record the water flow (used only visual comparison purposes).